# HIGH FREQUENCY MARKET MICROSTRUCTURE NOISE ESTIMATES AND LIQUIDITY MEASURES

By Yacine Aït-Sahalia[1,2] and Jialin Yu[2]

*Princeton University and Columbia University*

Using recent advances in the econometrics literature, we disentangle from high frequency observations on the transaction prices of a large sample of NYSE stocks a fundamental component and a microstructure noise component. We then relate these statistical measurements of market microstructure noise to observable characteristics of the underlying stocks and, in particular, to different financial measures of their liquidity. We find that more liquid stocks based on financial characteristics have lower noise and noise-to-signal ratio measured from their high frequency returns. We then examine whether there exists a common, market-wide, factor in high frequency stock-level measurements of noise, and whether that factor is priced in asset returns.

**1. Introduction.** Understanding volatility and its dynamics lies at the heart of asset pricing. As the primary measure of risk in modern finance, volatility drives the construction of optimal portfolios, the hedging and pricing of options and other derivative securities or the determination of a firm's exposure to a variety of risk factors and the compensation it can expect to earn from those risk exposures. It also plays a critical role in discovering trading and investment opportunities which provide an attractive risk-return trade-off.

It is therefore not surprising that volatility estimation and inference has attracted much attention in the financial econometric and statistical literature, including the seminal ARCH model of Engle (1982). Indeed, many estimators are available to measure an asset's volatility from a discrete price sample. But at least in the framework of parametric models, one will often start with the sum of squared log-returns, as not only the simplest and most

Received August 2007; revised February 2008.
[1]Supported in part by NSF Grants SES-03-50772 and DMS-05-32370.
[2]Supported by the Morgan Stanley equity market microstructure research grant.

natural estimator, but also as the one with the most desirable properties. For instance, in the context of parametric volatility models, this quantity will be not only the least squares estimator or the method of moments estimator with the sample variance as the moment function, but also the maximum likelihood estimator.

The asymptotic properties of this estimator are especially striking when sampling occurs at an increasing frequency which, when assets trade every few seconds, is a realistic approximation to what we observe using the now commonly available transaction or quote-level sources of financial data. In particular, as is well known in the context of stochastic processes, fully observing the sample path of an asset will in the limit perfectly reveal the volatility of that path. This result is nonparametric in nature, in that the estimator will converge to the quadratic variation of the process, a result which holds in great generality for semimartingales and does not rely on a parametric volatility model.

More recently, however, the statistical and econometric literatures have faced up to the fact that the situation in real life is not as simple as these asymptotic results suggest. Controlling for the market microstructure noise that is prevalent at high frequency has become a key issue. For a while, the approach used in the empirical literature consisted in ignoring the data sampled at the very highest frequencies out of concern for the noise that they might harbor, and sample instead once every 15 or 30 minutes.

The latest approach consists in explicitly incorporating microstructure noise into the analysis, and estimators have been developed to make use of all the data, no matter how high the frequency and how noisy, as prescribed by statistical principles. These methods make it possible to decompose the total observed variance into a component attributable to the fundamental price signal and one attributable to the market microstructure noise. These estimators can also produce consistent estimates of the magnitude of the market microstructure noise at high frequency, thereby producing a decomposition of total asset return volatility into fundamental and noise components.

Our objective in this paper is to better understand the nature of the information contained in these high frequency statistical measurements and relate them to observable financial characteristics of the underlying assets and, in particular, to different financial measures of their liquidity. Intuitively, one would expect that more liquid assets would tend to generate log-returns with a lower amount of microstructure noise, and a lower noise-to-signal ratio.

Market microstructure noise captures a variety of frictions inherent in the trading process: bid–ask bounces, discreteness of price changes, differences in trade sizes or informational content of price changes, gradual response of prices to a block trade, strategic component of the order flow, inventory control effects, etc. A better understanding of the relationship between these "micro" frictions and their "macro" consequences for asset prices' liquidity has implications for the asset management practice, especially for the strategies known as statistical arbitrage or proprietary trading.

This said, liquidity is an elusive concept. At a general level, the definition is straightforward: a market is liquid if one can trade a large quantity soon after wanting to do so, and at a price that is near the prices that precede and follow that transaction. How to translate that into an operative concept that is amenable to empirical analysis is less clear, and a variety of different measures have been proposed in the literature, including various measures of transaction costs, the extent to which prices depart from the random walk, etc.; see, for example, Amihud, Mendelson and Pedersen (2005) for a recent survey.

Our objective is therefore to examine the extent to which the high frequency statistical estimates that we will construct correlate with the various financial measures of liquidity, and whether they contain new or different information. In particular, we will look at whether high frequency estimates of microstructure noise contain a systematic, market-wide, risk factor and whether that risk factor is priced in the market, meaning that stocks that covary with our high-frequency measure of liquidity tend to get compensated in the form of higher returns. We will examine all these questions using a massive dataset consisting in all transactions recorded on all NYSE common stocks between June 1, 1995 and December 31, 2005.

The paper is organized as follows. In Section 2 we explain the strategies we use to estimate and separate the fundamental and noise volatilities. Section 3 describes our data. The empirical results where we relate the noise volatility to liquidity measures are in Section 4. In Section 5 we construct a semiparametric index model for the various financial measures of liquidity as they relate to our high frequency measurement: there, we construct the index of the diverse financial measures that best explains the statistical measurement of market microstructure noise. Then, we study in Section 6 whether there exists a common factor in stock-level liquidity measured at high frequency—we find that there is—and then in Section 7 whether that common factor is priced in asset returns—we find that the answer is yes, with some qualifications. Section 8 concludes.

**2. The noise and volatility estimators.** In this section we briefly review the two complementary estimation strategies that we will apply to decompose the returns' total variance into one due to the fundamental price and one due to market microstructure noise. The starting point to this analysis is a representation of the observed transaction log price at time $t$, $Y_t$, as the sum of an unobservable efficient price, $X_t$, and a noise component due to the imperfections of the trading process, $\varepsilon_t$:

$$(1) \qquad\qquad Y_t = X_t + \varepsilon_t.$$

One is often interested in estimating the volatility of the efficient log-price process $dX_t = \mu_t \, dt + \sigma_t \, dW_t$ using discretely sampled data on the transaction price process at times $0, \Delta, \ldots, n\Delta = T$.

The specification of the model coincides with that of Hasbrouck (1993), who interprets the standard deviation $a$ of the noise $\varepsilon$ as a summary measure of market quality. In Roll (1984), $\varepsilon$ is due entirely to the bid–ask spread, while Harris (1990b) lets additional phenomena give rise to $\varepsilon$. Examples include adverse selection effects as in Glosten (1987) and Glosten and Harris (1988); see also Madhavan, Richardson and Roomans (1997). A related literature has looked at transaction costs using bid–ask spread, price impact, etc., including Huang and Stoll (1996), Chan and Lakonishok (1997), and Cao, Choe and Hatheway (1997). When asymmetric information is involved, the disturbance $\varepsilon$ would typically no longer be uncorrelated with the process $W$ driving the efficient price and would also exhibit autocorrelation, which would complicate the analysis without fundamentally altering it; see the discussion below. Another important source of measurement error are rounding effects, since transaction prices are multiples of a tick size; see Gottlieb and Kalay (1985), Harris (1990a), Jacod (1996), and Delattre and Jacod (1997).

We will use below two classes of consistent estimators designed for the two situations where $\sigma_t$ is parametric (which can be reduced to $\sigma_t = \sigma$, a fixed parameter to be estimated), and $\sigma_t$ is nonparametric (i.e., an unrestricted stochastic process), in which case we seek to estimate the quadratic variation $\langle X, X \rangle_T = \int_0^T \sigma_t^2 \, dt$ over a fixed interval of time $T$, say, one day. In both cases, we are also interested in estimating consistently $a^2 = E[\varepsilon^2]$. For the parametric problem, we will use the maximum-likelihood estimator of Aït-Sahalia, Mykland and Zhang (2005a). For the nonparametric problem, we will use the estimator called Two Scales Realized Volatility of Zhang, Mykland and Aït-Sahalia (2005b), which is the first estimator shown to be consistent for $\langle X, X \rangle_T$.

The estimation of $\langle X, X \rangle_T$ has been studied in the constant $\sigma$ case by Zhou (1996), who proposes a bias correcting approach based on autocovariances. The behavior of this estimator has been studied by Zumbach, Corsi and Trapletti (2002). Hansen and Lunde (2006) study the Zhou estimator and extensions in the case where volatility is time varying but conditionally nonrandom. Related contributions have been made by Oomen (2006) and Bandi and Russell (2003). The Zhou estimator and its extensions, however, are inconsistent. This means in this particular case that, as the frequency of observation increases, the estimator diverges instead of converging to $\langle X, X \rangle_T$.

2.1. *The parametric case: constant volatility.* Consider first the parametric case studied in Aït-Sahalia, Mykland and Zhang (2005a), which by a change of variable and Itô's lemma can be immediately reduced to one where $\sigma$ is constant. If no market microstructure noise were present, that is, $\varepsilon \equiv 0$, the log-returns $R_i = Y_{\tau_i} - Y_{\tau_{i-1}}$ would be i.i.d. $N(0, \sigma^2 \Delta)$. The MLE for $\sigma^2$ then coincides with the realized volatility of the process, $\hat{\sigma}^2 = \frac{1}{T} \sum_{i=1}^{n} R_i^2$. Furthermore, $T^{1/2}(\hat{\sigma}^2 - \sigma^2) \underset{n \to \infty}{\longrightarrow} N(0, 2\sigma^4 \Delta)$ and, thus, selecting $\Delta$ as small as possible is optimal for the purpose of estimating $\sigma^2$.

When the observations are noisy, with the $\varepsilon's$ being i.i.d. noise with mean 0 and variance $a^2$, the true structure of the observed log-returns $R_i$ is given by an MA(1) process since $R_i = \sigma(W_{\tau_i} - W_{\tau_{i-1}}) + \varepsilon_{\tau_i} - \varepsilon_{\tau_{i-1}} \equiv u_i + \eta u_{i-1}$, where the $u's$ are mean zero and variance $\gamma^2$ with $\gamma^2(1 + \eta^2) = \mathrm{Var}[R_i] = \sigma^2 \Delta + 2a^2$ and $\gamma^2 \eta = \mathrm{Cov}(R_i, R_{i-1}) = -a^2$.

If we assume for a moment that $\varepsilon \sim N(0, a^2)$ (an assumption we will relax below), then the $u's$ are i.i.d. Gaussian and the likelihood function for the vector $R$ of observed log-returns, as a function of the transformed parameters $(\gamma^2, \eta)$, is given by

$$l(\eta, \gamma^2) = -\ln \det(V)/2 - n \ln(2\pi\gamma^2)/2 - (2\gamma^2)^{-1} R'V^{-1}R,$$

where

$$(2) \qquad V = [v_{ij}] = \begin{pmatrix} 1+\eta^2 & \eta & \cdots & 0 \\ \eta & 1+\eta^2 & \ddots & \vdots \\ \vdots & \ddots & \ddots & \eta \\ 0 & \cdots & \eta & 1+\eta^2 \end{pmatrix}.$$

Then the MLE $(\hat{\sigma}^2, \hat{a}^2)$ is consistent and its asymptotic variance is given by

$$\mathrm{AVAR}_{\mathrm{normal}}(\hat{\sigma}^2, \hat{a}^2) =$$
$$\begin{pmatrix} 4(\sigma^6 \Delta(4a^2 + \sigma^2 \Delta))^{1/2} + 2\sigma^4 \Delta & -\sigma^2 \Delta h \\ \bullet & \frac{\Delta}{2}(2a^2 + \sigma^2 \Delta)h \end{pmatrix}$$

with $h \equiv 2a^2 + (\sigma^2 \Delta(4a^2 + \sigma^2 \Delta))^{1/2} + \sigma^2 \Delta$.

Since $\mathrm{AVAR}_{\mathrm{normal}}(\hat{\sigma}^2)$ is increasing in $\Delta$, we are back to the situation where it is optimal to sample as often as possible. Interestingly, the AVAR structure of the estimator remains largely intact if we misspecify the distribution of the microstructure noise. Specifically, suppose that the $\varepsilon's$ have mean 0 and variance $a^2$ but are not normally distributed. If the econometrician (mistakenly) assumes that the $\varepsilon's$ are normal, inference is still done with the Gaussian log-likelihood $l(\sigma^2, a^2)$, using the scores $\dot{l}_{\sigma^2}$ and $\dot{l}_{a^2}$ as moment functions. Since the expected values of $\dot{l}_{\sigma^2}$ and $\dot{l}_{a^2}$ only depend on the second order moment structure of the log-returns $R$, which is unchanged by the absence of normality, the moment functions are unbiased: $E_{\mathrm{true}}[\dot{l}_{\sigma^2}] =$ $E_{\text{true}}[\dot{l}_{a^2}] = 0$ where "true" denotes the true distribution of the $Y's$. Hence, the estimator $(\hat{\sigma}^2, \hat{a}^2)$ based on these moment functions remains consistent and the effect of misspecification lies in the AVAR. By using the cumulants of the distribution of $\varepsilon$, we express the AVAR in terms of deviations from normality. We obtain that the estimator $(\hat{\sigma}^2, \hat{a}^2)$ is consistent and its asymptotic variance is given by

$$(3) \qquad \text{AVAR}_{\text{true}}(\hat{\sigma}^2, \hat{a}^2) = \text{AVAR}_{\text{normal}}(\hat{\sigma}^2, \hat{a}^2) + \text{Cum}_4[\varepsilon] \begin{pmatrix} 0 & 0 \\ 0 & \Delta \end{pmatrix},$$

where $\text{AVAR}_{\text{normal}}(\hat{\sigma}^2, \hat{a}^2)$ is the asymptotic variance in the case where the distribution of $U$ is Normal. $\varepsilon$ has mean zero, so in terms of its moments

$$(4) \qquad\qquad \text{Cum}_4[\varepsilon] = E[\varepsilon^4] - 3(E[\varepsilon^2])^2.$$

In the special case where $\varepsilon$ is normally distributed, $\text{Cum}_4[\varepsilon] = 0$.

The presence of a drift does not alter these earlier conclusions, not just because it would be economically irrelevant at the observation frequencies we consider, but also because of the following. Suppose that $X_t = \mu t + \sigma W_t$, then the block of the AVAR matrix corresponding to $(\hat{\sigma}^2, \hat{a}^2)$ is the same as if $\mu$ were known, in other words, as if $\mu = 0$, which is the case we focused on.

Aït-Sahalia, Mykland and Zhang (2005a) also discuss how the likelihood function is to be modified in the case of serially correlated noise and noise that is correlated with the price process. In those cases the form of the variance matrix of the observed log-returns must be altered, replacing $\gamma^2 v_{ij}$ with

$$\begin{aligned} \text{cov}(R_i, R_j) = {}&\sigma^2 \Delta \delta_{ij} + \text{cov}(\sigma(W_{\tau_i} - W_{\tau_{i-1}}), \varepsilon_{\tau_j} - \varepsilon_{\tau_{j-1}}) \\ &+ \text{cov}(\sigma(W_{\tau_j} - W_{\tau_{j-1}}), \varepsilon_{\tau_i} - \varepsilon_{\tau_{i-1}}) \\ &+ \text{cov}(\varepsilon_{\tau_i} - \varepsilon_{\tau_{i-1}}, \varepsilon_{\tau_j} - \varepsilon_{\tau_{j-1}}), \end{aligned}$$

where $\delta_{ij} = 1$ if $i = j$ and 0 otherwise. A model for the time series dependence of the $\varepsilon$ and its potential correlation to the price process would then specify the remaining terms.

2.2. *The nonparametric case: stochastic volatility.* An alternative model is nonparametric, where volatility is left unspecified, stochastic, and we now summarize the TSRV approach to separating the fundamental and noise volatilities in this case. When $dX_t = \sigma_t dW_t$, the object of interest is now the quadratic variation $\langle X, X \rangle_T = \int_0^T \sigma_t^2 \, dt$ over a fixed time period $[0, T]$. The usual estimator of $\langle X, X \rangle_T$ is the realized volatility (RV)

$$(5) \qquad\qquad [Y, Y]_T = \sum_{i=1}^{n} (Y_{t_{i+1}} - Y_{t_i})^2.$$

In the absence of noise, $[Y,Y]_T$ consistently estimates $\langle X, X \rangle_T$. The sum converges to the integral, with a known distribution, dating back to Jacod (1994) and Jacod and Protter (1998). As in the constant $\sigma$ case, selecting $\Delta$ as small as possible (i.e., $n$ as large as possible) is optimal.

But ignoring market microstructure noise leads to an even more dangerous situation than when $\sigma$ is constant and $T \to \infty$. After suitable scaling, RV based on the observed log-returns is a consistent and asymptotically normal estimator—but of the quantity $2nE[\varepsilon^2]$ rather than of the object of interest, $\langle X, X \rangle_T$. Said differently, in the high frequency limit, market microstructure noise totally swamps the variance of the price signal at the level of the realized volatility.

This is of course already visible in the special case of constant volatility. Since the expressions above are exact small-sample ones, they can, in particular, be specialized to analyze the situation where one samples at increasingly higher frequency ($\Delta \to 0$, say, sampled every minute) over a fixed time period ($T$ fixed, say, a day). With $N = T/\Delta$, we have

$$(6) \qquad E[\hat{\sigma}^2] = \frac{2na^2}{T} + o(n) = \frac{2nE[\varepsilon^2]}{T} + o(n),$$

$$(7) \qquad \mathrm{Var}[\hat{\sigma}^2] = \frac{2n(6a^4 + 2\,\mathrm{Cum}_4[\varepsilon])}{T^2} + o(n) = \frac{4nE[\varepsilon^4]}{T^2} + o(n),$$

so $(T/2n)\hat{\sigma}^2$ becomes an estimator of $E[\varepsilon^2] = a^2$ whose asymptotic variance is $E[\varepsilon^4]$. Note, in particular, that $\hat{\sigma}^2$ estimates the variance of the noise, which is essentially unrelated to the object of interest $\sigma^2$.

It has long been known that sampling as prescribed by $[Y,Y]_T^{(\mathrm{all})}$ is not a good idea. The recommendation in the literature has then been to sample sparsely at some lower frequency, by using a realized volatility estimator $[Y,Y]_T^{(\mathrm{sparse})}$ constructed by summing squared log-returns at some lower frequency: 5 mn, or 10, 15, 30 mn, typically [see, e.g., Andersen et al. (2001), Barndorff-Nielsen and Shephard (2002), and Gençay et al. (2002)]. Reducing the value of $n$, from say 23,400 (1 second sampling) to 78 (5 mn sampling over the same 6.5 hours), has the advantage of reducing the magnitude of the bias term $2nE[\varepsilon^2]$. Yet, one of the most basic lessons of statistics is that discarding data is, in general, not advisable.

Zhang, Mykland and Aït-Sahalia (2005b) proposed a solution to this problem which makes use of the full data sample, yet delivers consistent estimators of both $\langle X, X \rangle_T$ and $a^2$. The estimator, Two Scales Realized Volatility (TSRV), is based on subsampling, averaging, and bias-correction. By evaluating the quadratic variation at two different frequencies, averaging the results over the entire sampling, and taking a suitable linear combination of the result at the two frequencies, one obtains a consistent and asymptotically unbiased estimator of $\langle X, X \rangle_T$.

TSRV's construction is quite simple: first, partition the original grid of observation times, $G = \{t_0, \dots, t_n\}$ into subsamples, $G^{(k)}, k = 1, \dots, K$, where $n/K \to \infty$ as $n \to \infty$. For example, for $G^{(1)}$ start at the first observation and take an observation every 5 minutes; for $G^{(2)}$ start at the second observation and take an observation every 5 minutes, etc. Then we average the estimators obtained on the subsamples. To the extent that there is a benefit to subsampling, this benefit can now be retained, while the variation of the estimator will be lessened by the averaging. This reduction in the estimator's variability will open the door to the possibility of doing bias correction.

Averaging over the subsamples gives rise to the estimator

$$(8) \qquad [Y, Y]_T^{(\text{avg})} = \frac{1}{K} \sum_{k=1}^{K} [Y, Y]_T^{(k)}$$

constructed by averaging the estimators $[Y, Y]_T^{(k)}$ obtained on $K$ grids of average size $\bar{n} = n/K$. While a better estimator than $[Y, Y]_T^{(\text{all})}$, $[Y, Y]_T^{(\text{avg})}$ remains biased. The bias of $[Y, Y]_T^{(\text{avg})}$ is $2\bar{n}E[\varepsilon^2]$; of course, $\bar{n} < n$, so progress is being made. But one can go one step further. Indeed, $E[\varepsilon^2]$ can be consistently approximated using RV computed with all the observations:

$$(9) \qquad \widehat{E[\varepsilon^2]} = \frac{1}{2n} [Y, Y]_T^{(\text{all})}.$$

Hence, the bias of $[Y, Y]^{(\text{avg})}$ can be consistently estimated by $\bar{n}[Y, Y]_T^{(\text{all})}/n$. TSRV is the bias-adjusted estimator for $\langle X, X \rangle$ constructed as

$$(10) \qquad \widehat{\langle X, X \rangle}_T^{(\text{tsrv})} = \underbrace{[Y, Y]_T^{(\text{avg})}}_{\text{slow time scale}} - \frac{\bar{n}}{n} \underbrace{[Y, Y]_T^{(\text{all})}}_{\text{fast time scale}}.$$

If the number of subsamples is optimally selected as $K^* = cn^{2/3}$, then TSRV has the following distribution:

$$(11) \qquad \begin{aligned} \widehat{\langle X, X \rangle}_T^{(\text{tsrv})} &\overset{\mathcal{L}}{\approx} \underbrace{\langle X, X \rangle_T}_{\text{object of interest}} \\ &+ \frac{1}{n^{1/6}} \Big[ \underbrace{\frac{8}{c^2} E[\varepsilon^2]^2}_{\text{due to noise}} + \underbrace{c \frac{4T}{3} \int_0^T \sigma_t^4 \, dt}_{\text{due to discretization}} \Big]^{1/2} Z_{\text{total}}. \end{aligned}$$
$$\underbrace{\phantom{+ \frac{1}{n^{1/6}} \Big[ \frac{8}{c^2} E[\varepsilon^2]^2 + c \frac{4T}{3} \int_0^T \sigma_t^4 \, dt \Big]^{1/2} Z}}_{\text{total variance}}$$

Unlike all the previously considered ones, this estimator is now correctly centered.

A small sample refinement to $\widehat{\langle X, X \rangle}_T$ can be constructed as follows:

$$(12) \qquad \widehat{\langle X, X \rangle}_T^{(\text{tsrv,adj})} = \left( 1 - \frac{\bar{n}}{n} \right)^{-1} \widehat{\langle X, X \rangle}_T^{(\text{tsrv})}.$$

The difference with the estimator (10) is of order $O_p(K^{-1})$ and, thus, the two estimators have the same asymptotic behaviors to the order that we consider. However, the estimator (12) is unbiased to higher order.

So far, we have assumed that the noise $\varepsilon$ was i.i.d. In that case, log-returns are MA(1); it is possible to relax this assumption, and compute a TSRV estimator with two separate time scales [see Aït-Sahalia, Mykland and Zhang (2005b)]. TSRV provides the first consistent and asymptotic (mixed) normal estimator of the quadratic variation $\langle X, X \rangle_T$; as can be seen from (11), it has the rate of convergence $n^{-1/6}$. Zhang (2006) shows that it is possible to generalize TSRV to multiple time scales, by averaging not just on two time scales but on multiple time scales. For suitably selected weights, the resulting estimator, MSRV converges to $\langle X, X \rangle_T$ at the slightly faster rate $n^{-1/4}$. TSRV corresponds to the special case where one uses a single slow time scale in conjunction with the fast time scale to bias-correct it.

Finally, we exclude here any form of correlation between the noise $\varepsilon$ and the efficient price $X$, something which has been stressed by Hansen and Lunde (2006). As discussed in Aït-Sahalia, Mykland and Zhang (2006), however, the noise can only be distinguished from the efficient price under specific assumptions. In most cases, the assumption that the noise is stationary, alone, is not enough to make the noise identifiable. For example, coming back to the starting point (1) for the observed (log) price process $Y$, the model does not guarantee that one can always disentangle the signal or the volatility of the signal. To see this, suppose that the dynamics of the efficient price $X$ can be written as $dX_t = \mu_t \, dt + \sigma_t \, dW_t$, where the drift coefficient $\mu_t$ and the diffusion coefficient $\sigma_t$ can be random, and $W_t$ is a standard Brownian motion. If one assumed that the noise $\varepsilon_t$ is also an Itô process, say, $d\varepsilon_t = \nu_t \, dt + \gamma_t \, dB_t$, then $Y_t$ is also an Itô process of the form $dY_t = (\mu_t + \nu_t) \, dt + \omega_t \, dV_t$, where $\omega_t^2 = \sigma_t^2 + \gamma_t^2 + 2\sigma_t \gamma_t \, d\langle W, B \rangle_t / dt$. Unless one imposes additional constraints, it is therefore *not* possible to distinguish signal and noise in this model, and the only observable quadratic variation is $\int_0^T \omega_t^2 \, dt$, instead of the object of interest $\int_0^T \sigma_t^2 \, dt$.

Another issue we leave out is that of small sample corrections to the asymptotics of the estimators. Recently, Goncalves and Meddahi (2005) have developed an Edgeworth expansion for the basic RV estimator when there is no noise. Their expansion applies to the studentized statistic based on the standard RV and it is used for assessing the accuracy of the bootstrap in comparison to the first order asymptotic approach. By contrast, Zhang, Mykland and Aït-Sahalia (2005a) develop an Edgeworth expansion for nonstudentized statistics for the standard RV, TSRV, and other estimators, but allow for the presence of microstructure noise. Since we are using here point estimates for $a^2$ and $\langle X, X \rangle_T$, and the small sample corrections affect their distribution but not the point estimates, Edgeworth expansions are irrelevant to the present paper.

2.3. *Simulations: MLE or TSRV?*  We will implement below the ML and TSRV estimators on a large sample of NYSE stocks, consisting of all transactions recorded on all NYSE common stocks between June 1, 1995 and December 31, 2005. These stocks display a wide variety of characteristics. Many of them do not trade very frequently, especially at the beginning of the sample, to the point where some assumptions of the data generating process used in either the parametric or nonparametric models can be questioned: Is $\Delta$ small enough for the TSRV asymptotics to work? What is the impact of assuming that $\Delta$ is not random? Further, what is the impact of jumps in the price level and volatility, if any, on the MLE which assumes these effects away? What is the impact of stochastic volatility on the MLE? Relative to TSRV, to what extent does the efficiency of MLE outweigh its potential misspecification?

We now conduct Monte Carlo simulations, designed to be realistic given the nature of the data to which we will apply these estimators, to examine the impact of these various departures from the basic assumptions used to derive the properties of the estimators.[3]

It turns out that since we are estimating volatility and noise averages over a relatively short time interval $[0, T]$, where $T = 1$ day, assuming that the underlying values are constant over that time span is not adversely affecting the performance of the MLE of the average values of the underlying processes. Specifically, randomness in $\sigma_t$ over that time span, calibrated to multiples of the range of observed values, has little impact on the MLE. We will also see that the MLE is robust to incorporating a fair amount of jumps as well as randomness to the sampling intervals.

To see this, we perform simulations where the true data generating exhibits stochastic volatility:

$$\begin{align}
(13) \qquad dX_t &= \sqrt{V_t}\, dW_{1t}, \\
dV_t &= \kappa(v - V_t)\, dt + s\sqrt{V_t}\, dW_{2t},
\end{align}$$

where $W_{1t}$ and $W_{2t}$ are independent Brownian Motions. The parameters are $v = 0.1$ (corresponding to about 30% volatility per year) and $\kappa = 5$. $s$ is the volatility of the volatility parameter and will vary in our simulations, ranging from 0.1, 0.3, 0.5, 0.75, to 1. The choice of $\kappa = 5$ and $s$ around 0.5 is consistent with the estimates in Aït-Sahalia and Kimmel (2007). $V_0$ is initialized with its stationary distribution. The standard deviation of the noise, $a$, is set to 0.1%.

To add realism, we make the sampling interval $\Delta$ random; we assume an exponential distribution with mean $\overline{\Delta}$. By increasing $\overline{\Delta}$, we proxy for lower

---

liquidity in the sense of less active trading. We make the distribution of $\Delta$ independent of that of $X$; this is not completely realistic, but introducing a link between the two variables would change the likelihood function. With independence, we can treat the parameters of the distribution of $\Delta$ as nuisance parameters. 10,000 simulation sample paths are drawn. We run simulations for various combinations of the average sampling interval $\overline{\Delta}$ and the volatility of volatility parameter $s$.

The results are presented in Table 1 and Table 2. In these tables and the next, we report the average and the standard deviation (in parentheses) of MLE and TSRV estimates across the same 10,000 sample paths. The averages are to be compared to the true values $a^2 = 1\text{E-}6$ and $\sigma^2 = 0.1$, respectively. The sampling interval is random and exponentially distributed with mean $\overline{\Delta}$. TSRV is evaluated at $K = 25$ subsamples.

TABLE 1A

*Simulations: noise estimates from MLE under stochastic volatility with random sampling*

| MLE $\widehat{a^2}$ | $\overline{\Delta} = 1$ sec | $\overline{\Delta} = 5$ sec | $\overline{\Delta} = 10$ sec | $\overline{\Delta} = 30$ sec | $\overline{\Delta} = 2$ min | $\overline{\Delta} = 5$ min |
|---|---|---|---|---|---|---|
| $s = 0.1$ | 1.00E-6 | 1.00E-6 | 1.00E-6 | 1.00E-6 | 1.00E-6 | 1.00E-6 |
| | (1.30E-8) | (2.54E-8) | (3.73E-8) | (7.31E-8) | (1.93E-7) | (4.17E-7) |
| $s = 0.3$ | 1.00E-6 | 1.00E-6 | 1.00E-6 | 1.00E-6 | 1.00E-6 | 1.00E-6 |
| | (1.29E-8) | (2.53E-8) | (3.71E-8) | (7.28E-8) | (1.94E-7) | (4.13E-7) |
| $s = 0.5$ | 1.00E-6 | 1.00E-6 | 1.00E-6 | 1.00E-6 | 1.00E-6 | 1.00E-6 |
| | (1.29E-8) | (2.51E-8) | (3.69E-8) | (7.21E-8) | (1.95E-7) | (4.13E-7) |
| $s = 0.75$ | 1.00E-6 | 1.00E-6 | 1.00E-6 | 1.00E-6 | 1.00E-6 | 1.00E-6 |
| | (1.28E-8) | (2.52E-8) | (3.68E-8) | (7.31E-8) | (1.92E-7) | (4.13E-7) |
| $s = 1$ | 1.00E-6 | 1.00E-6 | 1.00E-6 | 1.00E-6 | 1.00E-6 | 1.00E-6 |
| | (1.29E-8) | (2.53E-8) | (3.72E-8) | (7.30E-8) | (1.93E-7) | (4.16E-7) |

TABLE 1B

*Simulations: noise estimates from TSRV under stochastic volatility with random sampling*

| TSRV $\widehat{a^2}$ | $\overline{\Delta} = 1$ sec | $\overline{\Delta} = 5$ sec | $\overline{\Delta} = 10$ sec | $\overline{\Delta} = 30$ sec |
|---|---|---|---|---|
| $s = 0.1$ | 1.01E-6 | 1.05E-6 | 1.09E-6 | 1.26E-6 |
| | (1.46E-8) | (2.73E-8) | (3.94E-8) | (7.56E-8) |
| $s = 0.3$ | 1.01E-6 | 1.05E-6 | 1.09E-6 | 1.26E-6 |
| | (1.46E-8) | (2.78E-8) | (4.05E-8) | (8.24E-8) |
| $s = 0.5$ | 1.01E-6 | 1.05E-6 | 1.09E-6 | 1.26E-6 |
| | (1.48E-8) | (2.89E-8) | (4.37E-8) | (9.57E-8) |
| $s = 0.75$ | 1.01E-6 | 1.05E-6 | 1.09E-6 | 1.26E-6 |
| | (1.51E-8) | (3.12E-8) | (4.87E-8) | (1.15E-7) |
| $s = 1$ | 1.01E-6 | 1.05E-6 | 1.09E-6 | 1.26E-6 |
| | (1.56E-8) | (3.43E-8) | (5.55E-8) | (1.38E-7) |

TABLE 2A
*Simulations: volatility estimates from MLE under stochastic volatility with random sampling*

| MLE $\widehat{\sigma^2}$ | $\overline{\Delta} = 1$ sec | $\overline{\Delta} = 5$ sec | $\overline{\Delta} = 10$ sec | $\overline{\Delta} = 30$ sec | $\overline{\Delta} = 2$ min | $\overline{\Delta} = 5$ min |
|---|---|---|---|---|---|---|
| $s = 0.1$ | 0.100 | 0.100 | 0.100 | 0.100 | 0.100 | 0.100 |
| | (0.007) | (0.009) | (0.011) | (0.014) | (0.021) | (0.028) |
| $s = 0.3$ | 0.100 | 0.100 | 0.100 | 0.100 | 0.100 | 0.100 |
| | (0.014) | (0.016) | (0.016) | (0.018) | (0.024) | (0.031) |
| $s = 0.5$ | 0.100 | 0.100 | 0.100 | 0.100 | 0.101 | 0.100 |
| | (0.023) | (0.024) | (0.024) | (0.026) | (0.030) | (0.036) |
| $s = 0.75$ | 0.100 | 0.100 | 0.100 | 0.101 | 0.100 | 0.100 |
| | (0.034) | (0.034) | (0.034) | (0.036) | (0.039) | (0.044) |
| $s = 1$ | 0.101 | 0.100 | 0.100 | 0.101 | 0.100 | 0.100 |
| | (0.045) | (0.045) | (0.045) | (0.047) | (0.048) | (0.052) |

TABLE 2B
*Simulations: volatility estimates from TSRV under stochastic volatility with random sampling*

| TSRV $\widehat{\sigma^2}$ | $\overline{\Delta} = 1$ sec | $\overline{\Delta} = 5$ sec | $\overline{\Delta} = 10$ sec | $\overline{\Delta} = 30$ sec |
|---|---|---|---|---|
| $s = 0.1$ | 0.100 | 0.100 | 0.099 | 0.097 |
| | (0.008) | (0.011) | (0.014) | (0.023) |
| $s = 0.3$ | 0.100 | 0.100 | 0.099 | 0.097 |
| | (0.015) | (0.017) | (0.019) | (0.025) |
| $s = 0.5$ | 0.100 | 0.099 | 0.099 | 0.097 |
| | (0.023) | (0.024) | (0.026) | (0.031) |
| $s = 0.75$ | 0.100 | 0.100 | 0.099 | 0.098 |
| | (0.034) | (0.034) | (0.035) | (0.040) |
| $s = 1$ | 0.101 | 0.099 | 0.099 | 0.097 |
| | (0.045) | (0.045) | (0.046) | (0.049) |

Next, we add jumps. The data generating process includes stochastic volatility and jumps in both level and volatility:

$$dX_t = \sqrt{V_t}\, dW_{1t} + J_t^X\, dN_{1t},$$

$$(14)$$

$$dV_t = \kappa(v - V_t)\, dt + s\sqrt{V_t}\, dW_{2t} + V_{t-}J_t^V\, dN_{2t}.$$

$N_{1t}$ and $N_{2t}$ are independent Poisson processes with arrival rate $\lambda_1$ and $\lambda_2$. In the simulations, we set for simplicity $\lambda_1 = \lambda_2 = \lambda$. The price jump size has the distribution $J_t^X \sim N(0, 0.02^2)$, that is, a one standard deviation jump changes the price level by 2%. The proportional jump size in volatility $J_t^V = \exp z$, where $z \sim N(-5, 1)$. As a result, the proportional jump size in volatility $J_t^V$ has a mean of 1 percent and a standard deviation of 1.5 percent. We fix $s = 0.5$. The other parameters are the same as those in (13). Simulations again incorporate a number of combinations of the sampling interval $\overline{\Delta}$ and jump intensity $\lambda$. The results are shown in Table 3 and Table 4. As in the previous tables, we report the average and the standard deviation (in parentheses) of MLE and TSRV estimates across the same 10,000 sample paths.

It can be seen from the results that in all cases the ML and TSRV estimators of $a^2$ are robust to various types of departures from the model's basic assumptions under a wide range of simulation design values, including properties of the volatility and the sampling mechanism. MLE assumes that volatility is nonstochastic; we find that for the purpose of applying the estimator over intervals of 1 day, any reasonable variability of volatility over that time span has no adverse effects on the estimator. Similarly, jumps and randomness in the sampling intervals, within a large range of values that contains the empirically relevant ones, do not affect the estimator.

TABLE 3A

*Simulations: noise estimates from MLE under stochastic volatility and jumps with random sampling*

| MLE $\widehat{a^2}$ | $\overline{\Delta} = 1$ sec | $\overline{\Delta} = 5$ sec | $\overline{\Delta} = 10$ sec | $\overline{\Delta} = 30$ sec | $\overline{\Delta} = 2$ min | $\overline{\Delta} = 5$ min |
|---|---|---|---|---|---|---|
| $\lambda = 4$ | 1.00E-6 | 1.00E-6 | 1.00E-6 | 1.00E-6 | 1.00E-6 | 1.00E-6 |
| | (1.27E-8) | (2.61E-8) | (3.89E-8) | (7.43E-8) | (1.96E-7) | (5.01E-7) |
| $\lambda = 12$ | 1.00E-6 | 1.00E-6 | 1.00E-6 | 1.00E-6 | 1.00E-6 | 1.00E-6 |
| | (1.27E-8) | (2.85E-8) | (4.07E-8) | (8.12E-8) | (2.11E-7) | (5.27E-7) |
| $\lambda = 52$ | 1.00E-6 | 1.00E-6 | 1.00E-6 | 1.00E-6 | 1.00E-6 | 1.02E-6 |
| | (1.35E-8) | (3.38E-8) | (5.40E-8) | (1.18E-7) | (4.20E-7) | (8.65E-7) |
| $\lambda = 252$ | 1.00E-6 | 1.00E-6 | 1.00E-6 | 1.00E-6 | 1.00E-6 | 1.07E-6 |
| | (1.62E-8) | (4.85E-8) | (8.47E-8) | (2.00E-7) | (6.55E-7) | (1.73E-6) |

TABLE 3B

*Simulations: noise estimates from TSRV under stochastic volatility and jumps with random sampling*

| TSRV $\widehat{a^2}$ | $\overline{\Delta} = 1$ sec | $\overline{\Delta} = 5$ sec | $\overline{\Delta} = 10$ sec | $\overline{\Delta} = 30$ sec |
|---|---|---|---|---|
| $\lambda = 4$ | 1.01E-6 | 1.05E-6 | 1.09E-6 | 1.26E-6 |
| | (1.50E-8) | (3.11E-8) | (5.09E-8) | (1.06E-7) |
| $\lambda = 12$ | 1.01E-6 | 1.05E-6 | 1.09E-6 | 1.27E-6 |
| | (1.54E-8) | (3.48E-8) | (5.54E-8) | (1.45E-7) |
| $\lambda = 52$ | 1.02E-6 | 1.06E-6 | 1.11E-6 | 1.32E-6 |
| | (1.83E-8) | (4.89E-8) | (8.50E-8) | (2.39E-7) |
| $\lambda = 252$ | 1.03E-6 | 1.09E-6 | 1.18E-6 | 1.52E-6 |
| | (2.74E-8) | (8.68E-8) | (1.68E-7) | (4.59E-7) |

TABLE 4A

*Simulations: volatility estimates from MLE under stochastic volatility and jumps with random sampling*

| MLE $\widehat{\sigma^2}$ | $\overline{\Delta} = 1$ sec | $\overline{\Delta} = 5$ sec | $\overline{\Delta} = 10$ sec | $\overline{\Delta} = 30$ sec | $\overline{\Delta} = 2$ min | $\overline{\Delta} = 5$ min |
|---|---|---|---|---|---|---|
| $\lambda = 4$ | 0.101 | 0.102 | 0.102 | 0.101 | 0.101 | 0.102 |
| | (0.030) | (0.031) | (0.039) | (0.030) | (0.038) | (0.042) |
| $\lambda = 12$ | 0.104 | 0.105 | 0.106 | 0.106 | 0.104 | 0.105 |
| | (0.041) | (0.046) | (0.047) | (0.051) | (0.044) | (0.056) |
| $\lambda = 52$ | 0.121 | 0.122 | 0.121 | 0.122 | 0.121 | 0.119 |
| | (0.086) | (0.088) | (0.089) | (0.094) | (0.096) | (0.087) |
| $\lambda = 252$ | 0.201 | 0.203 | 0.205 | 0.201 | 0.197 | 0.198 |
| | (0.174) | (0.180) | (0.189) | (0.183) | (0.189) | (0.190) |

TABLE 4B

*Simulations: volatility estimates from TSRV under stochastic volatility and jumps with random sampling*

| TSRV $\widehat{\sigma^2}$ | $\overline{\Delta} = 1$ sec | $\overline{\Delta} = 5$ sec | $\overline{\Delta} = 10$ sec | $\overline{\Delta} = 30$ sec |
|---|---|---|---|---|
| $\lambda = 4$ | 0.101 | 0.101 | 0.101 | 0.097 |
| | (0.030) | (0.031) | (0.038) | (0.035) |
| $\lambda = 12$ | 0.104 | 0.104 | 0.104 | 0.102 |
| | (0.041) | (0.046) | (0.046) | (0.053) |
| $\lambda = 52$ | 0.121 | 0.122 | 0.119 | 0.117 |
| | (0.086) | (0.087) | (0.085) | (0.089) |
| $\lambda = 252$ | 0.201 | 0.201 | 0.201 | 0.195 |
| | (0.173) | (0.175) | (0.182) | (0.176) |

TSRV is of course robust to stochastic volatility, but on the other hand, it is more sensitive to low sampling frequency, that is, high sampling intervals $\overline{\Delta}$, situations: the bias correction in TSRV relies on the idea that RV computed with all the data, $[Y, Y]_T^{(\text{all})}$, consists primarily of noise which is the notion that underlies (9). This is of course true asymptotically in $n$, that is, when $\overline{\Delta} \to 0$. But if the full data sample frequency is low to begin with, as, for instance, in the case of a stock sampled every minute instead of every second, $[Y, Y]_T^{(\text{all})}$ will not consist entirely of noise and bias-correcting on the basis of (9) may over-correct. Since these types of situations (low sampling frequency) will occur fairly often in our large sample below, the simulations argue for privileging MLE as the baseline estimator in our empirical application.

**3. The data.** We are now ready to examine the results produced by the estimators on real data and relate them to various financial measures of liquidity.

3.1. *High frequency stock returns.* We collect intra-day transaction prices and quotes from the NYSE Trade and Quote (TAQ) database, for all NYSE common stocks during the sample period of June 1, 1995 to December 31, 2005. Common stocks are defined as those in the Center for Research in Security Prices (CRSP) database whose SHRCD variable is either 10 or 11. The TAQ database starts in January 1993. Beginning in June 1995, the trade time in TAQ is the Consolidated Trade System (CTS) time stamp. Previously, the time shown was the time the trade information was received by the NYSE's Information Generation System, which is approximately 3 seconds later than the CTS time stamp.

3.2. *Liquidity measures.* We look at a wide collection of liquidity proxies. Two sets of liquidity measures are considered—a set of measures constructed from high frequency data (denoted as $H$) and a set of measures constructed from daily or lower frequencies (denoted as $D$).

We obtain daily share turnover, closing price, total number of shares outstanding, and monthly stock return from the Center for Research in Security Prices (CRSP) database. For stock $i$ in day $t$, let $\sigma_{i,t}$ denote the annualized stock return volatility, to be estimated as described in Section 2.1 from intra-day observations. We write $SPREAD_{i,t}$ for the average intra-day proportional bid–ask spread $(Ask - Bid)/Bid\,Ask\,Midpoint$. Only those intra-day observations with an ask price higher than the bid price are included. We let $LOGTRADESIZE_{i,t}$ denote the logarithm of the average number of shares per trade and $LOGNTRADE_{i,t}$ denote the log of the total number of intra-day trades. The vector $H$ of intra-day liquidity measures is

$$H = [\sigma, SPREAD, LOGTRADESIZE, LOGNTRADE]^T.$$

We let $LOGVOLUME_{i,t}$ be the log of daily share volume for stock $i$ on day $t$ obtained from the CRSP daily stock file. Let $MONTHVOL_{i,t}$ denote the annualized monthly stock return volatility for stock $i$ estimated using sixty monthly returns data in the most recent five-year window ending no later than $t$. Let $LOGP_{i,t}$ denote the log of stock $i$'s closing price on day $t$. We use $LOGSHROUT_{i,t}$ to denote the log of total shares outstanding for stock $i$ at the end of day $t$. These liquidity measures have been used to explain transaction costs in Huang and Stoll (1996), Chan and Lakonishok (1997), and Cao, Choe and Hatheway (1997).

Hasbrouck (2005) constructs a variety of annual liquidity measures. From Hasbrouck (2005), we obtain five liquidity measures: $cLogMean_{i,t}$ (Gibbs estimate of the log effective cost), $cMdmLogz_{i,t}$ (Moment estimate of the log effective cost, infeasible set to 0), $I2_{i,t}$ (square root variant of the Amihud illiquidity ratio), $L2_{i,t}$ (square root variant of liquidity ratio), $\gamma_{i,t}$ (Pastor and Stambaugh gamma). These measures are constructed annually for stock $i$ using observations in the most recent calendar year ending no later than $t$. We exclude those estimates constructed from less than sixty observations.

We also collected data on analyst coverage (from I/B/E/S database) and institutional ownership (from the CDA/Spectrum Institutional 13f Holdings database). Let $COVER_{i,t}$ denote the most recently reported number of analysts following stock $i$, and $LOGCOVER_{i,t} = \log(1 + COVER_{i,t})$. When a stock has no analyst coverage, $LOGCOVER_{i,t} = 0$. We use $IO_{i,t}$ to denote the most recently reported fraction of stock $i$'s total shares outstanding that are owned by institutions.

The vector $D$ of daily (or lower) frequency liquidity measures is

$$
\begin{aligned}
(15) \quad D = [ & LOGVOLUME, MONTHVOL, LOGP, cLogMean, \\
& cMdmLogz, I2, L2, \gamma, LOGSHROUT, LOGCOVER, IO]^T.
\end{aligned}
$$

The lower frequency measures ignore intra-day information, but have a longer time series available. The vector $A$ of all liquidity measures is

$$
(16) \qquad\qquad A = [H, D]^T.
$$

**4. The noise and volatility estimates.** We now relate the high frequency estimates of market microstructure noise to the financial measures of stock liquidity.

4.1. *High frequency estimates of microstructure noise and volatility.* Using log returns constructed from intra-day transaction prices, we estimate the market microstructure noise $a_{i,t}$ and the volatility $\sigma_{i,t}$ of stock $i$ on day $t$ using the MLE described in Section 2.1.[4] We exclude stock-day combinations with fewer than 200 intra-day transactions. Table 5 reports the basic summary statistics for the number of stocks and the daily number of high frequency observations. The average number of stocks in a typical day is 653. There are at least 61 stocks and at most 1278 stocks on any given day in the sample. There tend to be less stocks in the early part of the sample. The number of stocks varies also because, to be included in the sample, a stock-day combination is required to have a minimum of 200 intra-day transactions. There is an average of 910 transactions in a stock-day combination. The maximum number of intra-day transactions for one stock observed in this sample is 8445.

---

[4] The computer code in Matlab used in the estimation, together with the noise and volatility estimates for each stock-day combination, can be found in the supplementary material Aït-Sahalia and Yu (2009).

TABLE 5
*Summary statistics: daily stocks and trades 1995–2005*

|                                  | Mean | St. Er. | Min | Max   |
|----------------------------------|------|---------|-----|-------|
| Daily number of stocks           | 653  | 380     | 61  | 1,278 |
| Daily number of trades per stock | 910  | 808     | 200 | 8,445 |

TABLE 6
*MLE estimates of microstructure noise, fundamental volatility, and noise-to-signal ratio*

|                                        | Mean    | s.d.    |
|----------------------------------------|---------|---------|
| Noise $a_{j,t}$                        | 0.050%  | 0.050%  |
| Fundamental Volatility $\sigma_{j,t}$  | 34.8%   | 24.4%   |
| Noise-to-signal Ratio $NSR_{j,t}$      | 36.6%   | 19.4%   |

Table 6 reports the basic summary statistics for the noise and volatility estimates. Estimates for all stocks $j$ in all days $t$ of the sample period June 1, 1995–December 21, 2005 are pooled to compute the mean and standard deviation. The average noise standard deviation $a$ in the sample is 5 basis points (bps). The estimates of volatility $\sigma$ average to 34.8%. Figure 1 contains the histograms of MLE of $a$ and $\sigma$ estimated in our sample for all the stock-day combinations.

We will use the TSRV estimates as control variables for the MLE results. They are generally quite similar and do not produce economically meaningful differences. In order to save space, we will not report the corresponding results based on the TSRV estimates.

4.2. *Market microstructure noise and liquidity.* We begin by determining the extent to which our estimates of the market microstructure noise magnitude $a_{j,t}$ correlate with the liquidity measures that have been proposed in the literature. Specifically, for each liquidity measure $x$ in the vector $A$ in (16), we run the following regression:

$$(17) \qquad a_{j,t} = c_0 + x_{t-1}c_1 + \varepsilon_{i,t}.$$

The estimation results are in the first column of Table 7. This table reports the OLS regression results of market microstructure noise $a$ on individual liquidity measures one-by-one [column (1)], on all liquidity measures [column (2)], and on all those liquidity measures that can be constructed without using intra-day data [column (3)]. The noise $a$ and intra-day volatility $\sigma$ are estimated using maximum-likelihood estimation. The t-statistics are adjusted for heteroskedasticity and correlation within industry level using the Fama–French 48 industry classification [**Fama and French (1997)**].

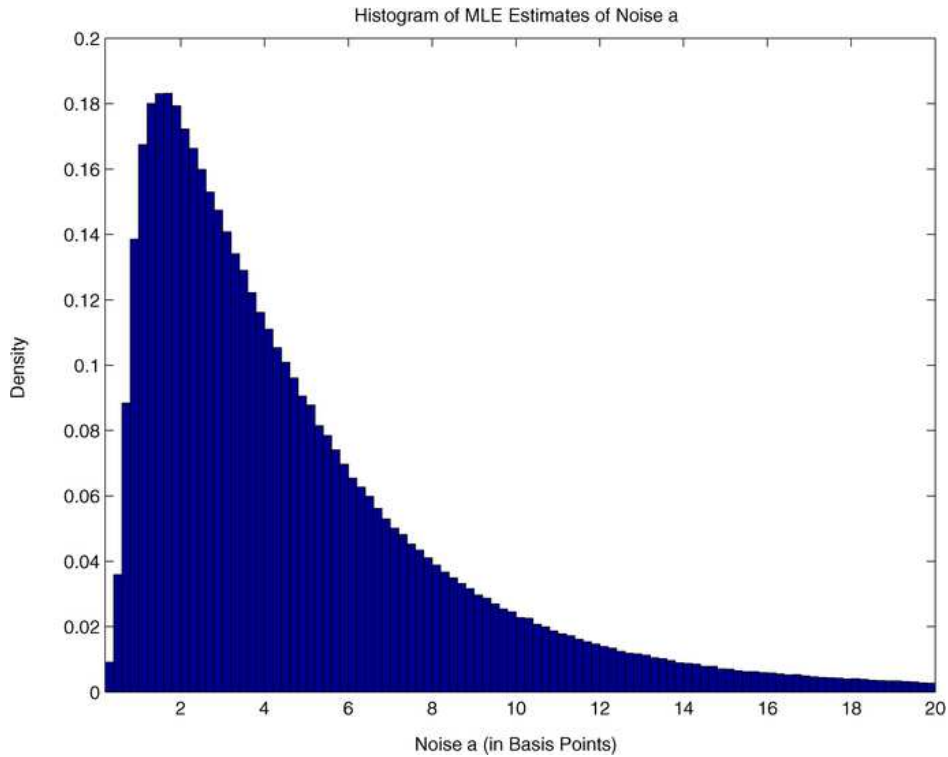

Fig. 1. *Distributions of MLE of microstructure noise and volatility, all NYSE stocks 1995–2005.*

The noise is positively correlated with volatility (both the intra-day and the monthly volatility), spread, transaction size, effective cost of trading, Amihud's Illiquidity ratio, and Pastor–Stambaugh's gamma. The noise is negatively correlated with number of intra-day transactions, price level, liquidity ratio, shares outstanding, analyst coverage, and institutional ownership. This is consistent with the notion that liquid stocks have less noise. The adjusted regression R-squared indicates that intra-day bid–ask spread explains most of the variation in noise (63%). Bid-ask bounces are a well-recognized phenomenon in transaction price data—indeed, the only source of noise in the model of Roll (1984). Among the daily liquidity measures, the price level explains the most variation in noise (28%).

We then look at the following two regressions:

$$(18) \qquad a_{j,t} = c_0 + A_{i,t-1}^T c_1 + \varepsilon_{i,t},$$

$$(19) \qquad a_{j,t} = c_0 + D_{i,t-1}^T c_1 + \varepsilon_{i,t},$$

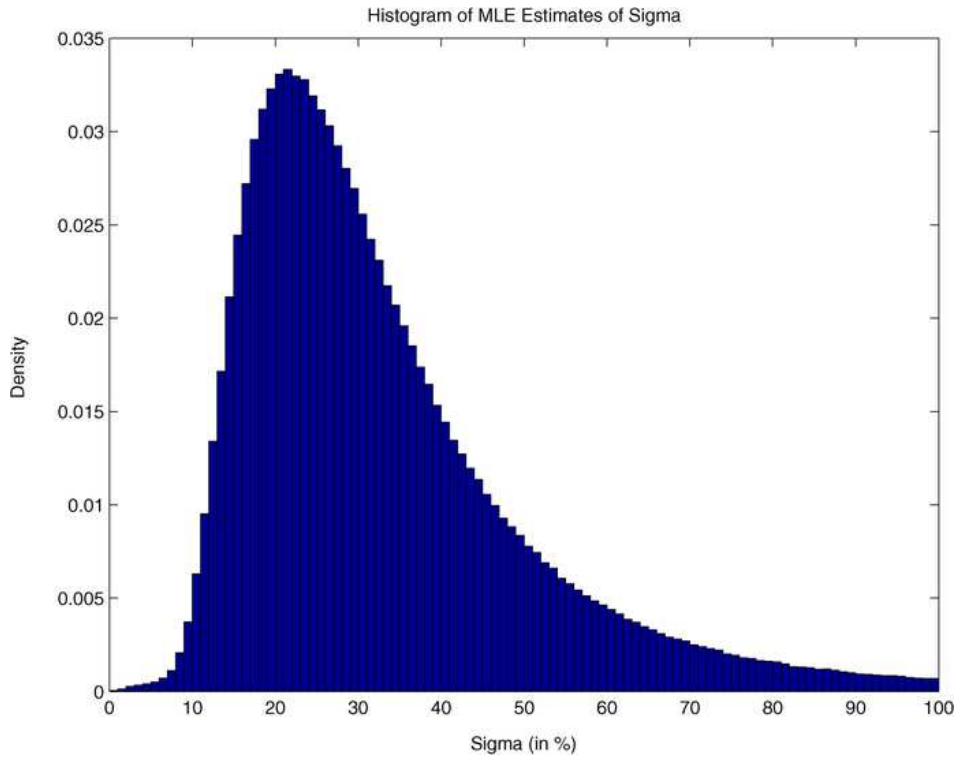

FIG. 1. *Continued.*

where the vector $A$ contains all liquidity measures in (16) and $D$ has all daily liquidity measures in (15) constructed without relying on intra-day observations.

The second column of Table 7 reports the regression results of (18). Intraday spread and price level are the most statistically significant explanatory variables, consistent with the result from (17). Some of the regression coefficients changed sign relative to the estimates of (17). This is not surprising since the explanatory variables are all correlated. The third column of the table reports the regression results of (19). The price level is now the most statistically significant regressor, which is not surprising given its impact on the bid–ask spread (a \$2 stock will not have the same spread as a \$200 stock with otherwise equivalent characteristics). Trading volume, which aggregates the information in trade size and number of trades, is positively correlated with noise.

4.3. *Noise-to-signal ratio and liquidity.* We use $NSR_{j,t}$ to denote the noise-to-signal ratio of stock $j$ on day $t$. When using MLE under the assumptions of Section 2.1, the proportion of the total return variance that is

TABLE 7
*Regression of market microstructure noise on liquidity measures*

| | (1) Individual measure | | | (2) All measures | | (3) Daily measures | |
|---|---|---|---|---|---|---|---|
| | Coef | t-stat | Adj $R^2$ | Coef | t-stat | Coef | t-stat |
| $\sigma$ | 0.0012 | 36.37 | 30.48% | −0.00025 | −8.46 | | |
| SPREAD | 0.17 | 33.77 | 63.44% | 0.17 | 32.68 | | |
| LOGTRADESIZE | 0.00020 | 15.86 | 12.31% | 2.9E−05 | 7.65 | | |
| LOGNTRADE | −0.00028 | −30.01 | 19.27% | −4.4E−05 | −6.04 | | |
| LOGVOLUME | | | | | | 0.000031 | 3.80 |
| MONTHVOL | 0.00063 | 10.73 | 4.93% | −3.0E−05 | −1.23 | −0.00055 | −4.68 |
| LOGP | −0.00039 | −26.76 | 27.57% | −0.00018 | −26.64 | −0.00043 | −16.80 |
| cLogMean | 0.062 | 14.10 | 7.00% | 0.0053 | 3.51 | 0.051 | 11.32 |
| cMdmLogz | 0.010 | 4.66 | 0.71% | −0.0015 | −1.84 | −0.020 | −10.11 |
| I2 | 8.9E−05 | 17.06 | 6.78% | −7.4E−06 | −3.37 | 2.5E−05 | 3.84 |
| L2 | −9.7E−05 | −16.25 | 7.05% | 1.3E−05 | 4.38 | −3.1E−05 | −4.04 |
| $\gamma$ | 8.1 | 1.65 | 0.03% | −0.29 | −0.35 | −0.27 | −0.08 |
| LOGSHROUT | −2.7E−05 | −3.14 | 0.40% | −7.4E−06 | −1.39 | −7.0E−05 | −3.95 |
| LOGCOVER | −8.0E−05 | −6.43 | 1.16% | 1.4E−05 | 2.37 | 0.00016 | 12.41 |
| IO | −0.00060 | −7.64 | 4.83% | −0.00014 | −5.54 | −0.00053 | −4.70 |
| Constant | | | | 0.0011 | 10.73 | 0.0030 | 6.89 |
| Adj$R^2$ | | | | 72.55% | | 37.09% | |

market microstructure-induced is

$$(20) \qquad NSR = \frac{2a^2}{\sigma^2 \Delta + 2a^2}$$

at observation interval $\Delta$. Here, $NSR$ is defined as a ratio of the noise variance to the total return variance, as opposed to the use of the term in other contexts to separate volatility from a trend.

As $\Delta$ gets smaller, $NSR$ gets closer to 1, so that a larger proportion of the variance in the observed log-return is driven by market microstructure frictions, and, correspondingly, a lesser fraction reflects the volatility of the underlying price process $X$. This effect is responsible for the divergence of traditional realized measures at high frequency: instead of converging to $\sigma^2$ as intended, they diverge according to $2na^2$, where $n = T/\Delta$ goes to infinity when $T = 1$ day is fixed and one samples at increasing frequency, $\Delta \to 0$.

Table 6 reports the summary statistics for the noise-to-signal ratio estimates constructed from estimates of $\sigma$ and $a$. Estimates for all stocks in all days of the sample period June 1, 1995–December 21, 2005 are pooled to compute the mean and standard deviation. The noise-to-signal ratio averages 36.6%.

We next examine the correlation of the noise-to-signal ratio $NSR_{j,t}$ with the existing liquidity measures contained in the vector $A$. Specifically, for

TABLE 8
*Regression of noise-to-signal ratio on liquidity measures*

| | (1) Individual measure | | | (2) All measures | | (3) Daily measures | |
|---|---|---|---|---|---|---|---|
| | Coef | t-stat | Adj$R^2$ | Coef | t-stat | Coef | t-stat |
| $\sigma$ | −0.025 | −2.60 | 0.10% | −0.28 | −18.26 | | |
| SPREAD | 16 | 14.37 | 3.61% | 21 | 11.31 | | |
| LOGTRADESIZE | 0.065 | 15.10 | 8.64% | 0.025 | 7.70 | | |
| LOGNTRADE | −0.023 | −3.95 | 0.84% | −0.023 | −5.60 | | |
| LOGVOLUME | | | | | | −0.017 | −5.43 |
| MONTHVOL | −0.0070 | −0.24 | 0.004% | −0.10 | −7.86 | −0.19 | −10.10 |
| LOGP | −0.096 | −19.75 | 11.13% | −0.12 | −23.65 | −0.13 | −30.03 |
| cLogMean | 5.4 | 3.24 | 0.36% | −3.8 | −3.50 | −1.6 | −1.31 |
| cMdmLogz | 2.2 | 3.52 | 0.23% | 1.3 | 2.81 | 0.38 | 0.67 |
| I2 | −0.0051 | −2.69 | 0.16% | −0.00018 | −0.15 | 0.00011 | 0.08 |
| L2 | 0.0080 | 2.41 | 0.34% | −0.00026 | −0.21 | −0.011 | −4.75 |
| $\gamma$ | 1823 | 2.86 | 0.01% | −1286 | −2.01 | −1249 | −2.50 |
| LOGSHROUT | 0.045 | 12.34 | 7.74% | 0.043 | 9.91 | 0.060 | 12.31 |
| LOGCOVER | 0.027 | 4.12 | 0.85% | 0.0077 | 1.64 | 0.028 | 5.82 |
| IO | −0.23 | −13.07 | 4.67% | −0.092 | −7.48 | −0.12 | −6.75 |
| Constant | | | | 0.11 | 1.33 | 0.046 | 0.56 |
| Adj$R^2$ | | | | 31.74% | | 25.15% | |

each liquidity measure $x$ in the vector $A$ in (16), we run the following regression:

$$(21) \qquad NSR_{j,t} = c_0 + x_{t-1}c_1 + \varepsilon_{i,t}.$$

The estimation results are reported in the first column of Table 8. As in the preceding table, column (1) reports the OLS regression results of noise-to-signal ratio $NSR$ on individual liquidity measures one-by-one, while column (2) includes on all liquidity measures, and column (3) all those liquidity measures that can be constructed without using intra-day data. Except for intra-day volatility, monthly volatility, illiquidity ratio, liquidity ratio, shares outstanding, and analyst coverage, the correlations between noise-to-signal ratio and liquidity measures have the same sign as the correlations between noise and liquidity measures. The negative correlation between noise-to-signal ratio $NSR$ and volatility is not surprising because the noise-to-signal ratio has volatility in the denominator. The positive correlations between $NSR$ and shares outstanding, analyst coverage, and liquidity ratio are likely due to the same reason, with more shares outstanding/analyst coverage/liquidity proxying less volatile stocks. The negative correlation between $NSR$ and the illiquidity ratio is likely due to the same reason, too. The price level explains the most variation in noise-to-signal ratio.

We then look at the following regressions:

$$(22) \qquad NSR_{j,t} = c_0 + A_{i,t-1}^T c_1 + \varepsilon_{i,t},$$

$$(23) \qquad NSR_{j,t} = c_0 + D_{i,t-1}^T c_1 + \varepsilon_{i,t}.$$

The second column of Table 8 reports the regression results of (22). The price level is the most statistically significant explanatory variables, consistent with the result from (21). The coefficients for most right-hand side variables have the same sign as those from (21). The coefficients for the Gibbs estimates of the effective trading cost, liquidity ratio, and Pastor–Stambaugh gamma changed sign. This is again not surprising since the explanatory variables are all correlated and some of these regressors are not significant to begin with. The third column of the table reports the regression results of (23). The price level remains the most significant regressor. Trading volume, which aggregates the information in trade size and number of trades, is negatively correlated with noise-to-signal ratio because trading volume is positively correlated with intra-day volatility.

### 4.4. *Robustness checks.*

#### 4.4.1. *Structural breaks in microstructure noise.*   We now proceed to check that the large $R^2$ obtained in the regressions reported in Sections 4.2 and 4.3 are not simply the product of structural breaks due to the two reductions of the tick size during the sample period. Indeed, regressing a variable with two level changes in the same direction on anything else with a trend (deterministic of stochastic) could produce a large $R^2$, which may or may not be spurious.

Structural breaks in the time series of the average noise magnitude $\bar{a}_t$ obtained from averaging the stock-level estimates of $a_{j,t}$ can be see in Figure 2: there are two main breaks, corresponding to the a reduction from a 1/8 to a 1/16 minimum tick size on June 24, 1997; the second corresponds to the move to decimalization, that is, a further reduction in the tick size to 1/100 on January 29, 2001. Since bid–ask bounces represent the leading cause of market microstructure noise in our transactions-based price data, it is natural to check that these two breaks, leading to decreases in the average value of $\bar{a}_t$, are not by themselves giving rise to spurious results.

For that purpose, we re-do the regressions in column (1) of Tables 7 and 8, except that we also include two dummy variables, *SIXTEENTH* and *DECIMAL*, in the regressions isolating the three sample periods where minimum tick sizes were, respectively, 1/8, 1/16, and 1/100. *SIXTEENTH* equals one in the sample period where minimum tick size is 1/16 and zero otherwise. *DECIMAL* equals one in the sample period where minimum tick size is 1/100 and zero otherwise. The results are largely similar to those in Tables 7 and 8 and are omitted for brevity. The adjusted $R^2$ in these regressions are, if anything, slightly higher than those in Tables 7 and 8 because the two dummy variables capture some variations in noise.

### 4.4.2. *Fixed effect regression.*

In addition to the OLS regressions (17), (18), and (19), we also run the following stock fixed-effect regression to account for firm-specific heterogeneity. Specifically, for each individual liquidity measure $x$, we run

$$a_{j,t} = c_0 + x_{t-1}c_1 + STOCK\ FIXED\ EFFECT + \varepsilon_{i,t}.$$

For the vector $A$ of all liquidity measures and for the vector $D$ of liquidity measures constructed without relying on intra-day observations, we run

$$a_{j,t} = A^T_{i,t-1}c + STOCK\ FIXED\ EFFECT + \varepsilon_{i,t},$$

$$a_{j,t} = D^T_{i,t-1}c + STOCK\ FIXED\ EFFECT + \varepsilon_{i,t}.$$

A similar set of stock fixed effect regressions are run for *NSR* to double check the results from (21)–(23).

The results from the fixed effect regressions are similar to those from the OLS regressions. For brevity, the results are omitted and can be obtained from the authors upon request.

### 4.4.3. *Nonlinearities.*

We re-run the noise-to-signal ratio regressions (22) by including $\sigma^2$ in addition to $\sigma$ as explanatory variable to account for potential nonlinearity. Similarly, for the noise regression (18), we re-run it with the following permutations: replace the dependent variable noise $a$ with $a^2$, and include $\sigma^2$ in addition to $\sigma$ as explanatory variable. The results are similar to what we found previously. For brevity, the results are not reported here but are available from the authors upon request.

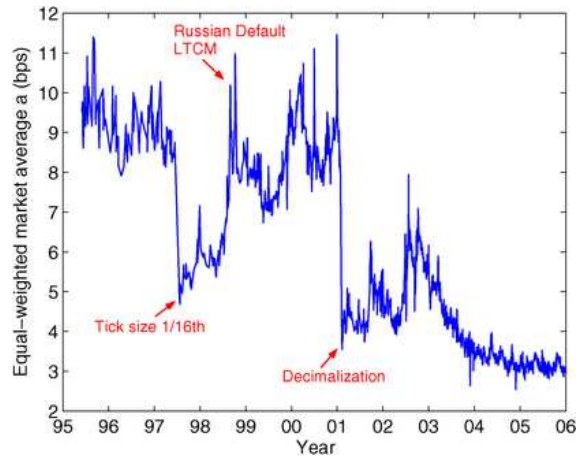

FIG. 2. *Daily equal-weighted average of microstructure noise a.*

The results are consistent with the conclusions in Section 5, namely, that a linear combination of the liquidity measures in (18), (19), (22), and (23) can accurately capture the variation in noise and noise-to-signal ratio.

**5. Semiparametric index of microstructure noise.** There are many different financial measures of liquidity contained in the vector $A$. We would like to be able to construct a single index that parsimoniously captures the variation in noise or noise-to-signal ratio using the various financial measures of liquidity that have been proposed in the market-microstructure literature. By summarizing the multidimensional vector of liquidity measures $A$ into an (endogenously determined) index $A^T b$, we reduce the sampling error associated with the various measures and allow for more robust out-of-sample extrapolation and forecasting.

We will estimate a semiparametric single index model

$$(24) \qquad y = E[y|A] = m(A) = g(A^T b),$$

where $y$ is either noise $a$ or noise-to-signal ratio $NSR$, $A$ is the vector of liquidity measures, $b$ is an unknown vector of coefficients, and $g(\cdot)$ is the unknown functional form linking $y$ and the index $A^T b$. The restriction imposed by the index structure is that $y$ depends on $A$ only through its potentially nonlinear dependence on the single index $A^T b$. The specific way in which the index averages the various measures of liquidity is not pre-specified – it will be estimated. We use the classical index model method to estimate $g(\cdot)$ and $b$; see, for example, Härdle and Linton (1994) for details on the semi-parametric single index model and the related estimation methods.[5]

Figure 3 plots the estimated semi-parametric link function $\widehat{g}(\cdot)$ for the noise against the single index $A^T\widehat{\delta}$ constructed using, respectively, all liquidity measures and daily liquidity measures. Similarly, Figure 4 plots the estimated semi-parametric link function $\widehat{g}(\cdot)$ for the noise-to-signal ratio constructed using, respectively, all liquidity measures and daily liquidity measures. The single index $A^T\widehat{\delta}$ is standardized to have zero mean and standard deviation of one. Also plotted in Figures 3 and 4 are the various quantiles for the estimates of noise $a$ and the noise-to-signal ratio, respectively.

As seen in both figures, the variations in noise or noise-to-signal ratio can be adequately captured by the single index $A^T\widehat{\delta}$, and a linear link function $g(\cdot)$ approximates fairly well the unknown functional form $m(\cdot)$. Importantly, the link functions are all increasing, which is consistent with a positive dependence between our statistical, high frequency, estimates of liquidity and

---

[5]We use a standard Gaussian kernel in the estimation and the bandwidths are chosen using Silverman's rule [see Härdle and Linton (1994)].

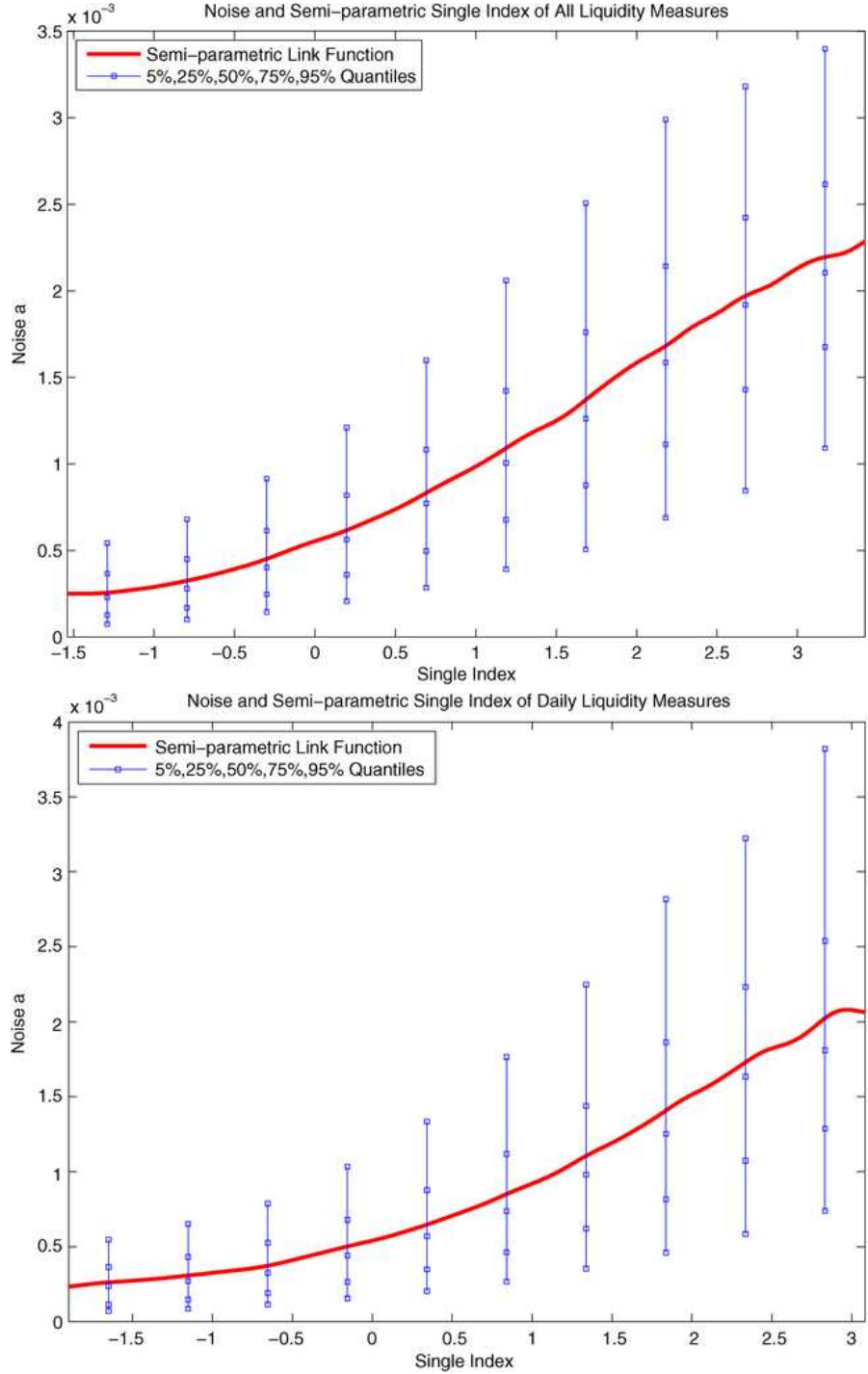

FIG. 3.  *Semi-parametric link function for the noise estimates.*

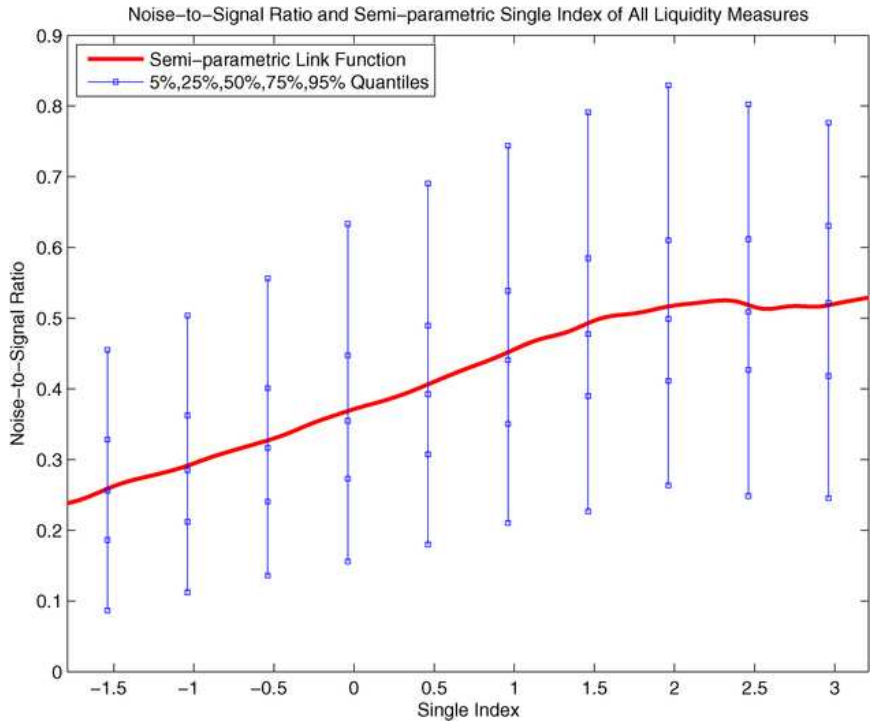

Fɪɢ. 4.   *Semi-parametric link function for the noise-to-signal ratio estimates.*

the financial liquidity measures documented in the sections above. Based on this, we will construct a single index from linear regressions on asset pricing implications, which will not rely on nonparametric estimation of the unknown link function $g(\cdot)$. Rather, we will work under the restriction that $g(\cdot)$ is linear, using the construction of the index of the various liquidity measures provided by the estimates of the index coefficient vector $b$.

Besides reducing the dimensionality of the regression, which in theory has robustness advantages, one further advantage of using the single index from daily liquidity measures in asset pricing is that it allows extrapolation, making it possible to use a longer time series of otherwise unavailable financial liquidity measures for some individual stocks.

**6. Market-wide liquidity risk.**   There are many reasons to expect that liquidity across many different stocks could co-move. First, at the microstructure level, factors believed to affect the provision of liquidity are subject to common factors: for instance, dealers' adjustments to their inventory levels in response to price or volatility movements (which we know are partly co-movements) can lead to adjustments to individual bid–ask spreads, quoted depths, and other liquidity proxies that then exhibit co-variation across wide

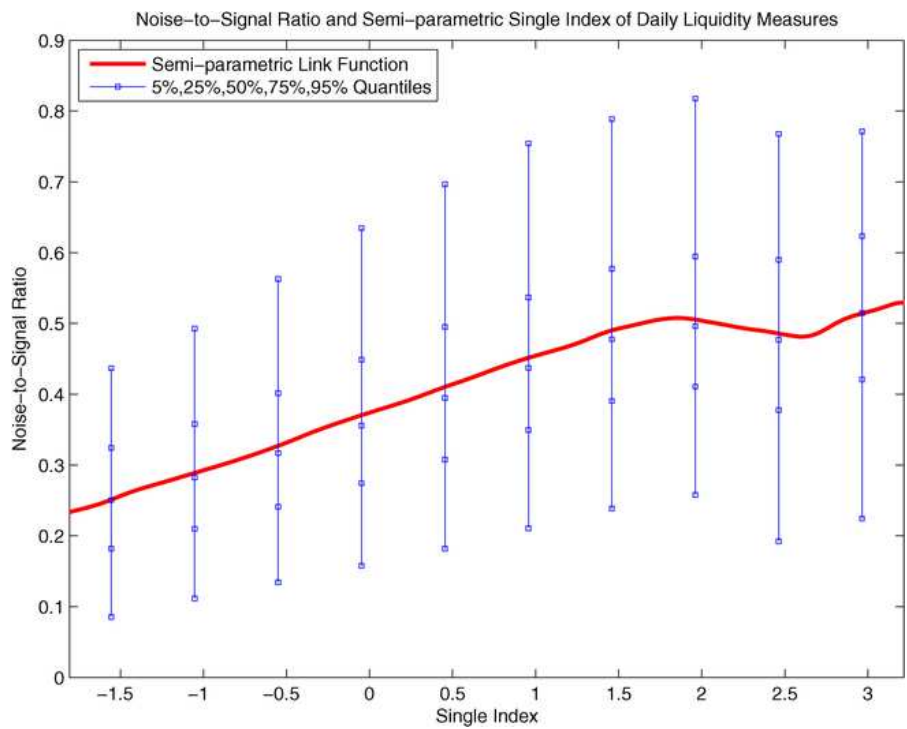

Fɪɢ. 4.   *Continued.*

segments of the market. Second, at the macroeconomic level, periods of un-
certainty are generally accompanied by a market-wide reduction in trading
activity as investors sit on the sidelines waiting for the uncertainty to get
resolved. Similarly, shifts in the perception of an asymmetric information
risk can also lead to co-movements in liquidity, perhaps on a more limited
scale (say, industry-wide.)[6]

---

[6]Chordia, Roll and Subrahmanyam (2000) examine whether quoted spreads, quoted
depth, and effective spreads co-move with market- and industry-wide liquidity. After
controlling for individual liquidity characteristics such as volatility, volume, and price,
they find that these co-movements remain significant. Chordia, Roll and Subrahmanyam
(2001) find that daily changes in market averages of liquidity and trading activity are
time-varying and negatively autocorrelated. When stock returns fall, so does liquidity.
Periods of volatility are followed by a decrease in trading activity. Finally, they document
day-of-the-week patterns, with Fridays experiencing lower trading activity and liquidity.
Hasbrouck and Seppi (2001) examine common movements in various liquidity proxies,
such as the bid–ask spread and quote sizes, which have relatively small common factors.
Huberman and Halka (2001) estimate models for quotes and depths for portfolios and find
evidence of common factors in liquidity in the form of residuals being correlated across
portfolios.

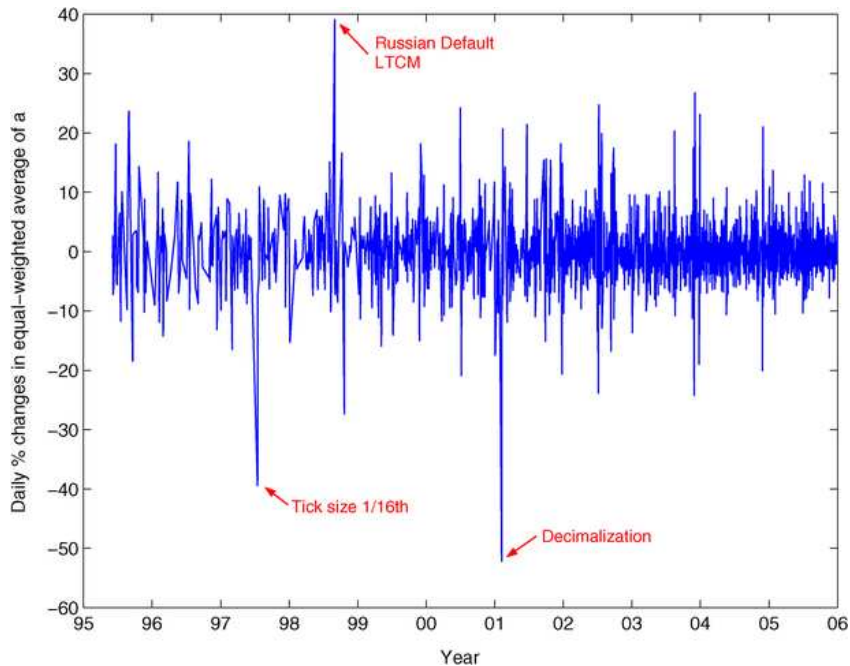

Fig. 5.   *Daily percentage changes of the equal-weighted average of microstructure noise a.*

We now examine whether similar findings hold for our measure of liquidity, based on intraday high frequency returns. In particular, we examine whether the time series of stock-level liquidity measures we have constructed above is subject to common factor variation, or whether it is primarily driven by microeconomic events and firm-specific variation, such as news announcements.

We start by looking at market-wide liquidity, as measured by an equal-weighted cross-sectional average of the estimates $a$ we constructed above. Figure 5 plots the daily percentage variation in market liquidity for the sample period of June 1995–December 2005. The aggregate liquidity can have sizeable fluctuations from time to time. Higher $a$ in the plot indicates low liquidity. The biggest upward spike takes place in the summer of 1998 when the liquidity is widely perceived to have dried up due to Russian default and LTCM collapse. The biggest downward spike occurs in February 2001 when NYSE went decimal. There is a similar downward spike around mid-1997 when NYSE reduced tick size from one-eighth to one-sixteenth. We have also constructed market liquidity using the value-weighted average of $a$ and the plot looks similar. The pairwise correlation between changes in equal-weighted and value-weighted averages of $a$ is 0.963 in this sample period.

TABLE 9
*Commonality in liquidity*

| | (1) | (2) | | (3) |
|---|---|---|---|---|
| | **Market** | **Market** | **Industry** | **Market** |
| Concurrent | 1.027 | 1.553 | 0.188 | 0.944 |
| t-stat | (1.95) | (1.99) | (0.56) | (2.12) |
| % positive | 62.8% | 57.1% | 52.3% | 56.8% |
| % + significant | 12.6% | 10.7% | 7.2% | 9.1% |
| Lag | | | | 0.353 |
| t-stat | | | | (1.28) |
| % positive | | | | 54.2% |
| % + significant | | | | 7.3% |
| Lead | | | | 0.348 |
| t-stat | | | | (1.54) |
| % positive | | | | 49.8% |
| % + significant | | | | 5.7% |
| Sum | | | | 1.645 |
| t-stat | | | | (2.42) |
| Median | 0.472 | 0.380 | 0.059 | 0.319 |
| p-value | 0.000 | 0.000 | 0.035 | 0.000 |
| Mean Adj$R^2$ | 0.003 | | 0.002 | 0.007 |
| Median Adj$R^2$ | 0.000 | | 0.001 | 0.002 |

We then look at the empirical covariation between individual stock liquidity (as measured by our stock-level estimate of $a$) and market-wide liquidity (as measured by the equal-weighted average of the $a_t$ estimates). We do this by regressing daily percentage changes in our individual stock liquidity measure $a_{j,t}$ on a constant and the daily percentage changes in the market-wide liquidity measure $a_{M,t}$:

$$(25) \qquad \ln(a_{j,t}/a_{j,t-1}) = \alpha_j + \beta_j \ln(a_{M,t}/a_{M,t-1}) + \eta_{j,t}.$$

We remove each individual stock from the computation of the market-wide liquidity average used in that stock's regression, so that the right-hand side regressor does not contain the left-hand side variable, and the estimated coefficients in all those regressions are not artificially constrained.

The results are reported in column (1) of Table 9. Column (1) of this table conducts, for each stock, a time-series regression of the daily log changes in individual stock liquidity measure $a$ on log changes in the equal-weighted cross-sectional average $a$ for all stocks in the sample ("Market"). The cross-sectional average of time series slope coefficients is reported with t-statistics in the parenthesis. "% positive" reports the percentage of positive slope coefficients, while "% + significant" gives the percentage of time-series regression t-statistics [from Newey and West (1987)] that are greater than 1.645 (the 5% critical level in a one-sided test). Column (2) reports the results from a time-series regression of daily log changes in individual stock $a$ on log changes in both equal-weighted market average of $a$ and equal-weighted industry average of $a$ ("Industry"). Industry is classified using one-digit SIC codes. Slope coefficients for both the market and the industry are reported. Column (3) conducts a time-series regression of daily log changes in individual stock $a$ on log changes in market-wide $a$ in the current, the previous, and the next trading days. In the table "sum" = concurrent + lag + lead slope coefficients, "median" is the median of time-series slope coefficients in columns (1) and (2), or the median of "Sum" in column (3), and "p-value" is the p-value of a signed test of the null hypothesis that median = 0. Cross-sectional mean and median adjusted R-squared of the time-series regressions are also reported. The liquidity measure $a$ of an individual stock is excluded from the construction of the market and the industry liquidity averages used in that stock's time-series regression.

The cross-sectional average of $\beta_j$ is 1.027 with a t-statistic of 1.95. 62.8% of the slope coefficients are positive. For each time-series regression, we obtain the Newey and West (1987) t-statistic for the slope coefficient. 12.6% of the time-series regression t-statistics are greater than 1.645, the 5% critical level in a one-sided test for the positivity of the slope coefficient. The cross-sectional median of the slope coefficients is 0.472 and the null hypothesis of a zero median is rejected with p-value = 0.000 in favor of a positive slope coefficient. There is a large unexplained component of individual stock liquidity variations, as is clear from the low regression adjusted $R^2$. This is consistent with the findings in Chordia, Roll and Subrahmanyam (2000) and Hasbrouck and Seppi (2001).

We have also run the corresponding regressions with value-weighted instead of equal-weighted market-wide liquidity averages and find similar results. For example, the cross-sectional average of $\beta_j$ is 2.376 with a t-statistic of 4.53 when a value-weighted average of market liquidity is used. These results suggest that there is a common component in individual stock liquidities measured by $a_{j,t}$.

We next investigate common industry components in stock-level liquidity $a$ by regressing daily percentage changes in our individual stock liquidity measure $a_{j,t}$ on a constant, the daily percentage changes in the market-wide liquidity measure $a_{M,t}$, and the daily percentage changes in the industry-level liquidity measure $a_{I,t}$:

$$\ln(a_{j,t}/a_{j,t-1}) = \alpha_j + \beta_{j,M} \ln(a_{M,t}/a_{M,t-1}) + \beta_{j,I} \ln(a_{I,t}/a_{I,t-1}) + \eta_{j,t}.$$

$a_{I,t}$ is constructed from the equal-weighted industry average of individual stock $a$. Firm $j$ is excluded from the construction of the industry-level liquidity used in its own regression. One-digit SIC code is used as industry classification, though we have also used Fama–French 5 industry and Fama–French 10 industry classifications and obtained similar results. Column (2) of Table 9 reports the regression results. The slope coefficient for market-wide liquidity is 1.553, similar to that in column (1), and remains statistically significant (t-statistic = 1.99). The industry slope coefficient is positive. However, both the economic and statistical significance is much smaller compared to that of the market coefficient. This is suggestive of some co-movements in industry liquidity, though the effect is measured with some noise. We have re-run the regression using value-weighted market and industry average liquidity and found similar results.

As a robustness check, we have run the regression specification (25) with one lag and one lead of changes in market liquidity. The lag and the lead are intended to capture delayed responses to common determinants of liquidity. The result is in column (3) of Table 9. The concurrent slope coefficient is 0.944, similar to those in columns (1) and (2). It is statistically significant with a t-statistic of 2.12. Both the lag and the lead coefficients are positive, though both their economic and statistical significance is smaller than the concurrent coefficient. The sum of the concurrent, lag, and lead coefficients averages to 1.645 with a t-statistic of 2.42, confirming that individual liquidity measures $a$ do co-move to some degree with each other.

**7. Asset pricing implications.** Given the evidence above that there is a common factor in liquidity as measured by our high frequency estimates, we now ask whether that common factor is priced. We begin by looking at the returns of portfolios sorted on various liquidity measures, including our high frequency measurement of the magnitude of the market microstructure noise, $a$. At the end of June in each year we sort stocks into quintiles using one of the liquidity measures. Monthly value-weighted portfolio returns are calculated for the twelve months following the sort. Table 10 shows the time-series averages of the monthly portfolio returns in the sample period of July 1995–June 2005. For the noise $a$, the annual sort at the end of June is based on the average of daily MLE estimates of $a$ within June. To reduce the estimation variability, we require at least a week's worth of $a$ estimates (i.e., at least five daily estimates) to construct the monthly average. The results are similar if different minimum numbers of daily estimates are used. We construct monthly liquidity measures for $NSR$, $\sigma$, $SPREAD$, $LOGTRADESIZE$, $LOGNTRADE$, $LOGVOLUME$. For share price $LOGP$ and total number of shares outstanding $LOGSHROUT$, we use the information from the last trading day of June to construct the monthly measure.

Table 10 shows the portfolio returns sorted using all NYSE stocks. The returns are monotonically increasing for portfolios sorted on $a$. The portfolio corresponding to the highest quintile $a$ outperforms the portfolio with the lowest quintile $a$ by 44 basis points (bps) per month (5.3% per year), and the difference is statistically significant. The portfolio returns sorted on $NSR$ are not monotonic across the five $NSR$ quintiles due to the effect of $\sigma$ on returns, though the portfolio with the lowest $NSR$ underperforms the other four quintiles. Portfolio returns sorted on another common liquidity measure, $SPREAD$, are roughly monotonic, although the return difference between the top and bottom quintiles is only about 20 bps per month. There are no clear return implications for the other liquidity measures.

TABLE 10
*Monthly portfolio returns of all NYSE stocks in quintiles sorted by different liquidity measures*

|  | **Low** | **2** | **3** | **4** | **High** |
|---|---|---|---|---|---|
| Intra-day measures |  |  |  |  |  |
| $a$ | 0.93 | 1.05 | 1.20 | 1.28 | 1.37 |
| $NSR$ | 0.65 | 0.98 | 1.07 | 1.15 | 1.00 |
| $\sigma$ | 1.09 | 0.90 | 1.03 | 0.94 | 0.84 |
| $SPREAD$ | 1.00 | 1.13 | 1.17 | 1.22 | 1.20 |
| $LOGTRADESIZE$ | 1.16 | 1.43 | 1.29 | 1.14 | 0.54 |
| $LOGNTRADE$ | 1.23 | 1.05 | 1.30 | 1.41 | 0.95 |
| Lower-frequency measures |  |  |  |  |  |
| $LOGVOLUME$ | 1.21 | 1.27 | 1.43 | 1.38 | 0.83 |
| $MONTHVOL$ | 0.99 | 1.00 | 1.10 | 1.14 | 1.09 |
| $LOGP$ | 1.10 | 1.07 | 1.12 | 1.05 | 0.91 |
| $cLogMean$ | 1.11 | 1.13 | 0.69 | 1.14 | 0.87 |
| $cMdmLogz$ | 1.09 | 1.04 | 0.86 | 1.18 | 0.67 |
| $I2$ | 0.94 | 1.29 | 1.14 | 1.43 | 1.03 |
| $L2$ | 1.08 | 1.39 | 1.15 | 1.29 | 0.95 |
| $\gamma$ | 1.31 | 1.23 | 0.94 | 1.10 | 1.32 |
| $LOGSHROUT$ | 1.02 | 1.23 | 1.42 | 1.18 | 0.94 |
| $LOGCOVER$ | 1.27 | 0.80 | 1.19 | 1.04 | 1.02 |
| $IO$ | 0.82 | 1.04 | 1.15 | 1.10 | 0.93 |

To check the effect of low-price stocks, Table 11 shows the portfolio returns sorted using all NYSE stocks whose price at the end of June is at least $5. The result is similar.

It is important to note from Tables 10 and 11 that *none* of the other liquidity measures in the vector $A$ appears to be priced the way $a$ is. This provides evidence that $a$ as a measure of liquidity contains information that is not captured by any single one of the traditional financial measures of market microstructure noise and liquidity. To obtain a liquidity factor with similar pricing power as $a$, one needs to group all the liquidity measures into the index $A^T b$ constructed above. At that point, we have an effective proxy for $a$.

To see whether the extra return earned by the illiquid portfolio is compensation for risk, we run, for each given $i = 1, 2, \ldots, 5$, the following time-series

TABLE 11
*Monthly liquidity-sorted portfolio returns of all NYSE stocks with price at least 5 dollars*

|  | **Low** | **2** | **3** | **4** | **High** |
|---|---|---|---|---|---|
| Intra-day measures |  |  |  |  |  |
| $a$ | 0.93 | 1.06 | 1.19 | 1.27 | 1.31 |
| $NSR$ | 0.66 | 0.96 | 1.08 | 1.19 | 0.99 |
| $\sigma$ | 1.09 | 0.93 | 1.00 | 1.00 | 0.72 |
| $SPREAD$ | 1.00 | 1.17 | 1.10 | 1.22 | 1.15 |
| $LOGTRADESIZE$ | 1.16 | 1.44 | 1.30 | 1.16 | 0.55 |
| $LOGNTRADE$ | 1.21 | 1.05 | 1.31 | 1.41 | 0.94 |
| Lower-frequency measures |  |  |  |  |  |
| $LOGVOLUME$ | 1.22 | 1.27 | 1.40 | 1.39 | 0.83 |
| $MONTHVOL$ | 0.99 | 1.00 | 1.08 | 1.21 | 1.03 |
| $LOGP$ | 1.12 | 1.05 | 1.10 | 1.05 | 0.90 |
| $cLogMean$ | 1.10 | 1.15 | 0.68 | 1.18 | 0.80 |
| $cMdmLogz$ | 1.09 | 1.03 | 0.95 | 1.18 | 0.67 |
| $I2$ | 0.94 | 1.29 | 1.14 | 1.43 | 1.01 |
| $L2$ | 1.06 | 1.38 | 1.16 | 1.30 | 0.95 |
| $\gamma$ | 1.29 | 1.24 | 0.95 | 1.05 | 1.32 |
| $LOGSHROUT$ | 1.03 | 1.22 | 1.42 | 1.18 | 0.94 |
| $LOGCOVER$ | 1.29 | 0.79 | 1.19 | 0.97 | 1.03 |
| $IO$ | 0.82 | 1.04 | 1.18 | 1.10 | 0.93 |

TABLE 12
*Monthly liquidity-sorted portfolio alphas of all NYSE stocks*

|  | **Low** | **2** | **3** | **4** | **High** |
|---|---|---|---|---|---|
| Intra-day measures |  |  |  |  |  |
| $a$ | 0.13 | 0.25 | 0.36 | 0.52 | 0.44 |
| $NSR$ | −0.32 | 0.14 | 0.25 | 0.39 | 0.24 |
| $\sigma$ | 0.42 | 0.12 | 0.17 | −0.08 | −0.47 |
| $SPREAD$ | 0.22 | 0.31 | 0.36 | 0.37 | 0.22 |
| $LOGTRADESIZE$ | 0.37 | 0.67 | 0.49 | 0.36 | −0.27 |
| $LOGNTRADE$ | 0.53 | 0.32 | 0.57 | 0.67 | 0.10 |
| Lower-frequency measures |  |  |  |  |  |
| $LOGVOLUME$ | 0.50 | 0.53 | 0.68 | 0.63 | −0.01 |
| $MONTHVOL$ | 0.34 | 0.31 | 0.25 | 0.10 | −0.22 |
| $LOGP$ | 0.21 | 0.30 | 0.31 | 0.26 | 0.10 |
| $cLogMean$ | 0.42 | 0.39 | −0.12 | 0.22 | −0.13 |
| $cMdmLogz$ | 0.25 | −0.28 | 0.23 | 0.38 | −0.20 |
| $I2$ | 0.15 | 0.51 | 0.36 | 0.62 | 0.08 |
| $L2$ | 0.17 | 0.58 | 0.39 | 0.52 | 0.14 |
| $\gamma$ | 0.42 | 0.39 | 0.14 | 0.36 | 0.60 |
| $LOGSHROUT$ | 0.13 | 0.43 | 0.59 | 0.39 | 0.15 |
| $LOGCOVER$ | 0.34 | −0.09 | 0.41 | 0.28 | 0.22 |
| $IO$ | 0.05 | 0.26 | 0.33 | 0.23 | 0.13 |

regression:

$$r_{i,t} - r_{f,t} = \alpha_i + \beta_i \cdot (r_{M,t} - r_{f,t}) + \varepsilon_{i,t},$$

where $i$ indicates one of the liquidity-sorted quintile portfolios. $r_{i,t}$ is the portfolio return in month $t$. $r_{f,t}$ is the one-month Treasury-bill rate. $r_{M,t}$ is the stock market return. The estimate of $\beta_i$ measures the exposure to the market risk. The estimate of $\alpha_i$ (CAPM alpha) measures the return unexplained by the exposure to the market factor which is then attributed to the sort on liquidity. Table 12 reports the CAPM alpha for the quintile portfolios constructed from all NYSE stocks. The monthly alpha of the top quintile $a$-stocks is 31 bps (3.7% per year) higher than the bottom quintile stocks, about 30% smaller than the raw stock return difference. The alphas across the five quintile portfolios don't sum up to zero because the quintile portfolios contain only NYSE stocks, yet the market portfolio is measured by the commonly used CRSP value-weighted return of all NYSE/AMEX/NASDAQ stocks. The spread between top and bottom *NSR* quintile stocks increased to 56 bps. This is partly due to the return spread across $\sigma$-quintiles [see Ang et al. (2006)]. *SPREAD* does not correlate with return alpha. We have also calculated CAPM alphas for liquidity-sorted portfolios constructed from a restricted sample of all NYSE stocks whose end of June price is at least $5. The results are similar and omitted for brevity.

In addition to the CAPM alpha, we also calculated alpha relative to the Fama–French 3-factor model [see Fama and French (1993)]. Specifically, we run, for each given $i = 1, 2, \ldots, 5$, the following time-series regression:

$$r_{i,t} - r_{f,t} = \alpha_i + \beta_i^M \cdot (r_{M,t} - r_{f,t}) + \beta_i^{HML} \cdot r_{HML,t} + \beta_i^{SMB} \cdot r_{SMB,t} + \varepsilon_{i,t},$$

where $r_{HML,t}$ and $r_{SMB,t}$ are the returns of two portfolios constructed to mimic risk factors associated with value and size, respectively.[7] The estimate of $\alpha_i$ (Fama–French 3-factor alpha) measures the return unexplained by exposure to market, value-, and size-related risks. When considering alphas with respect to the Fama–French 3-factor included, we find that higher $a$ portfolios of liquidity-sorted stocks no longer have significantly higher returns alphas. This suggests that the extent to which this liquidity factor is priced in the marketplace is either too weak to be observable in the sample when high frequency data are available after controlling by a multifactor asset pricing model or, alternatively, that the additional two factors in the Fama–French model proxy to some extent for the liquidity as measured by these estimates.[8]

---

[7]The portfolios, which are constructed and subsequently rebalanced at the end of each June, are based on the intersections of 2 portfolios formed on size (market equity, ME) and 3 portfolios formed on the ratio of book equity to market equity (BE/ME). The size breakpoint for year t is the median NYSE market equity at the end of June of year t. BE/ME for June of year t is the book equity for the last fiscal year end in t-1 divided by ME for December of t-1. The BE/ME breakpoints are the 30th and 70th NYSE percentiles. Growth/neutral/value stocks refer to those stocks with the lowest 30%/middle 40%/highest 30% BE/ME. *SMB* (Small Minus Big) is the average return on the three small portfolios minus the average return on the three big portfolios, $SMB = 1/3$ (Small Value + Small Neutral + Small Growth) $-1/3$ (Big Value + Big Neutral + Big Growth). HML (High Minus Low) is the average return on the two value portfolios minus the average return on the two growth portfolios, HML $= 1/2$ (Small Value + Big Value) $-1/2$ (Small Growth + Big Growth). See Fama and French (1993) for more details on the SMB and the HML factors.

**8. Conclusions.**   In this paper we decomposed the transaction prices of NYSE stocks into a fundamental component and a microstructure noise component. We relate the two components to observable financial characteristics and, in particular, to different observable measures of stock liquidity. We find that less noise, as measured statistically by lower estimates of the magnitude of the noise at high frequency, correlates positively with financial measures of liquidity, either at the daily frequency or at higher frequencies. More liquid stocks have lower noise and noise-to-signal ratio. Using daily liquidity measures, we construct a single index of the various financial measures of liquidity. We find that there is a common factor in liquidity as measured by our high frequency estimates, and that this common factor is priced by the market.

**Acknowledgments.**   We are very grateful for the comments of the Editor, an Associate Editor, and two referees.

SUPPLEMENTARY MATERIAL

---

[8]Brennan and Subrahmanyam (2002) study the relationship between monthly returns and measures of illiquidity obtained from intraday returns. They find that liquidity is significant for asset returns, even after accounting for the Fama–French factors (market return, size, value) as well as the stock price level. Pastor and Stambaugh (2003) also find that marketwide liquidity risk, measured at the daily frequency, is a state variable important for asset pricing: in their sample, the average return on stocks with high loadings on the liquidity factor is 7.5% higher than on those with low loadings, after adjusting for exposures to the three Fama–French factors plus momentum.

## REFERENCES

AÏT-SAHALIA, Y. and KIMMEL, R. (2007). Maximum likelihood estimation of stochastic volatility models. *Journal of Financial Economics* **83** 413–452.

AÏT-SAHALIA, Y., MYKLAND, P. A. and ZHANG, L. (2005a). How often to sample a continuous-time process in the presence of market microstructure noise. *Review of Financial Studies* **18** 351–416.

AÏT-SAHALIA, Y., MYKLAND, P. A. and ZHANG, L. (2005b). Ultra high frequency volatility estimation with dependent microstructure noise. Discussion paper, Princeton Univ.

AÏT-SAHALIA, Y., MYKLAND, P. A. and ZHANG, L. (2006). Comments on "Realized variance and market microstructure noise." *J. Bus. Econom. Statist.* **24** 162–167.

AÏT-SAHALIA, Y. and YU, J. (2009). Supplement to "High frequency market microstructure noise estimates and liquidity measures." DOI: 10.1214/08-AOAS200SUPP.

AMIHUD, Y., MENDELSON, H. and PEDERSEN, L. H. (2005). Liquidity and asset prices. *Foundations and Trends in Finance* **1** 269–364.

ANDERSEN, T. G., BOLLERSLEV, T., DIEBOLD, F. X. and LABYS, P. (2001). The distribution of exchange rate realized volatility. *J. Amer. Statist. Assoc.* **96** 42–55. MR1952727

ANG, A., HODRICK, R. J., XING, Y. and ZHANG, X. (2006). The cross-section of volatility and expected returns. *Journal of Finance* **51** 259–299.

BANDI, F. M. and RUSSELL, J. R. (2003). Microstructure noise, realized volatility and optimal sampling. Discussion paper, Univ. Chicago Graduate School of Business.

BARNDORFF-NIELSEN, O. E. and SHEPHARD, N. (2002). Econometric analysis of realized volatility and its use in estimating stochastic volatility models. *J. Roy. Statist. Soc. Ser. B* **64** 253–280. MR1904704

BRENNAN, M. J. and SUBRAHMANYAM, A. (2002). Market microstructure and asset pricing: On the compensation for illiquidity in stock returns. *Journal of Financial Economics* **41** 441–464.

CAO, C., CHOE, H. and HATHEWAY, F. (1997). Does the specialist matter? Differential execution costs and intersecurity subsidization on the New York stock exchange. *Journal of Finance* **52** 1615–1640.

CHAN, L. and LAKONISHOK, J. (1997). Institutional equity trading costs: NYSE versus Nasdaq. *Journal of Finance* **52** 713–735.

CHORDIA, T., ROLL, R. and SUBRAHMANYAM, A. (2000). Commonality in liquidity. *Journal of Financial Economics* **56** 3–28.

CHORDIA, T., ROLL, R. and SUBRAHMANYAM, A. (2001). Market liquidity and trading activity. *Journal of Finance* **56** 501–530.

DELATTRE, S. and JACOD, J. (1997). A central limit theorem for normalized functions of the increments of a diffusion process, in the presence of round-off errors. *Bernoulli* **3** 1–28. MR1466543

ENGLE, R. F. (1982). Autoregressive conditional heteroskedasticity with estimates of the variance of U.K. inflation. *Econometrica* **50** 987–1007. MR0666121

FAMA, E. F. and FRENCH, K. R. (1993). Common risk factors in the returns on stocks and bonds. *Journal of Financial Economics* **33** 3–56.

FAMA, E. F. and FRENCH, K. R. (1997). Industry costs of equity. *Journal of Financial Economics* **43** 143–193.

GENÇAY, R., BALLOCCHI, G., DACOROGNA, M., OLSEN, R. and PICTET, O. (2002). Real-time trading models and the statistical properties of foreign exchange rates. *Internat. Econom. Rev.* **43** 463–491.

GLOSTEN, L. R. (1987). Components of the bid–ask spread and the statistical properties of transaction prices. *Journal of Finance* **42** 1293–1307. MR0919477

GLOSTEN, L. R. and HARRIS, L. E. (1988). Estimating the components of the bid/ask spread. *Journal of Financial Economics* **21** 123–142.

GONCALVES, S. and MEDDAHI, N. (2005). Bootstrapping realized volatility. Discussion paper, Univ. Montréal.

GOTTLIEB, G. and KALAY, A. (1985). Implications of the discreteness of observed stock prices. *Journal of Finance* **40** 135–153.

HANSEN, P. R. and LUNDE, A. (2006). Realized variance and market microstructure noise. *J. Bus. Econom. Statist.* **24** 127–161.

HÄRDLE, W. and LINTON, O. (1994). Applied nonparametric methods. In *Handbook of Econometrics* (R. F. Engle and D. L. McFadden, eds.) **4** 2295–2339. Elsevier, Amsterdam. MR1315973

HARRIS, L. (1990a). Estimation of stock price variances and serial covariances from discrete observations. *Journal of Financial and Quantitative Analysis* **25** 291–306.

HARRIS, L. (1990b). Statistical properties of the roll serial covariance bid/ask spread estimator. *Journal of Finance* **45** 579–590.

HASBROUCK, J. (1993). Assessing the quality of a security market: A new approach to transaction-cost measurement. *Review of Financial Studies* **6** 191–212.

HASBROUCK, J. (2005). Trading costs and returns for US equities: Evidence from daily data. Discussion paper, New York Univ.

HASBROUCK, J. and SEPPI, D. J. (2001). Common factors in prices, order flows, and liquidity. *Journal of Financial Economics* **59** 383–411.

HUANG, R. and STOLL, H. (1996). Dealer versus auction markets: A paired comparison of execution costs on NASDAQ and the NYSE. *Journal of Financial Economics* **41** 313–357.

HUBERMAN, G. and HALKA, D. (2001). Systematic liquidity. *Journal of Financial Research* **24** 161–178.

JACOD, J. (1994). Limit of random measures associated with the increments of a Brownian semimartingale. Discussion paper, Univ. Paris-6.

JACOD, J. (1996). La Variation Quadratique du Brownien en Présence d'Erreurs d'Arrondi. *Astérisque* **236** 155–162. MR1417980

JACOD, J. and PROTTER, P. (1998). Asymptotic error distributions for the Euler method for stochastic differential equations. *Ann. Probab.* **26** 267–307. MR1617049

MADHAVAN, A., RICHARDSON, M. and ROOMANS, M. (1997). Why do security prices change? *Review of Financial Studies* **10** 1035–1064.

NEWEY, W. K. and WEST, K. D. (1987). A simple, positive semi-definite, heteroskedasticity and autocorrelation consistent covariance matrix. *Econometrica* **55** 703–708. MR0890864

OOMEN, R. C. (2006). Properties of realized variance under alternative sampling schemes. *J. Bus. Econom. Statist.* **24** 219–237. MR2234448

PASTOR, L. and STAMBAUGH, R. F. (2003). Liquidity risk and expected stock returns. *Journal of Political Economy* **111** 642–685.

ROLL, R. (1984). A simple model of the implicit bid–ask spread in an efficient market. *Journal of Finance* **39** 1127–1139.

ZHANG, L. (2006). Efficient estimation of stochastic volatility using noisy observations: A multi-scale approach. *Bernoulli* **12** 1019–1043. MR2274854

ZHANG, L., MYKLAND, P. A. and Y. AÏT-SAHALIA (2005a). Edgeworth expansions for realized volatility and related estimators. Discussion paper, Princeton Univ.

ZHANG, L., MYKLAND, P. A. and Y. AÏT-SAHALIA (2005b). A tale of two time scales: Determining integrated volatility with noisy high-frequency data. *J. Amer. Statist. Assoc.* **100** 1394–1411.

Zhou, B. (1996). High-frequency data and volatility in foreign-exchange rates. *J. Bus. & Econom. Statist.* **14** 45–52.

Zumbach, G., Corsi, F. and Trapletti, A. (2002). Efficient estimation of volatility using high frequency data. Discussion paper, Olsen & Associates.

Bendheim Center for Finance
Princeton University
26 Prospect Avenue
Princeton, New Jersey 08540-5296
USA
E-mail: yacine@princeton.edu

Finance and Economic Division
Columbia University
421 Uris Hall
3022 Broadway
New York, New York 10027
USA
E-mail: jy2167@columbia.edu